\documentclass[12pt]{article}
\usepackage{graphicx}
\usepackage{titlesec}
    \titleformat{\section}{\normalfont\Large\filcenter}{\thesection}{1em}{\MakeUppercase}

\usepackage[margin=1.0in]{geometry}
\usepackage{color}
\usepackage{times}
\usepackage{amsmath}
\usepackage{amsfonts}
\usepackage{lscape}

\makeatletter
\renewcommand\@biblabel[1]{}
\renewenvironment{thebibliography}[1]
{ \section*{\refname}
  \@mkboth{\MakeUppercase\refname}{\MakeUppercase\refname}
    \list{}
      {%
        \leftmargin0pt
        \advance\leftmargin\bibindent
        \itemindent -\bibindent
        \usecounter{enumiv}%
        \let\p@enumiv\@empty
        \baselineskip=12pt
      }%
      }
\makeatother

\begin{document}

\title{Bayesian low rank and sparse covariance matrix decomposition}
\author{Lin Zhang, Abhra Sarkar , Bani K. Mallick\\
Department of Statistics, Texas A\&M University }
\date{\vspace{-5ex}}

\maketitle

\begin{center}
\textbf{Abstract}
\end{center}

We consider the problem of estimating high-dimensional covariance matrices of a particular structure, which is a summation of low rank and sparse matrices. This covariance structure has a wide range of applications including factor analysis and random effects models. We propose a Bayesian method of estimating the covariance matrices by representing the covariance model in the form of a factor model with unknown number of latent factors. We introduce binary indicators for factor selection and rank estimation for the low rank component combined with a Bayesian lasso method for the sparse component estimation. Simulation studies show that our method can recover the rank as well as the sparsity of the two components respectively. We further extend our method to a graphical factor model where the graphical model of the residuals as well as selecting the number of factors is of interest. We employ a hyper-inverse Wishart prior for modeling decomposable graphs of the residuals, and a Bayesian graphical lasso selection method for unrestricted graphs. We show through simulations that the extended models can recover both the number of latent factors and the graphical model of the residuals successfully when the sample size is sufficient relative to the dimension.\\

\vspace*{.3in}


\newpage
\section{Introduction}

Estimation of covariance matrices is a fundamental issue in multivariate analysis with many statistical applications including modeling genetic data, brain imaging data, climate data, and many other fields. Suppose $\mathbf{y}_1, \ldots, \mathbf{y}_n$ are $q$-dimensional random vectors which independently follow a  multivariate Gaussian distribution $\mathcal{N}_q(\boldsymbol{\mu},\Sigma)$. It is well known that the sample covariance $\hat{\Sigma} = \sum^n_{i=1}(\mathbf{y}_i-\bar{\mathbf{y}})(\mathbf{y}_i-\bar{\mathbf{y}})'/(n-1)$ is not a stable estimator of the population covariance matrix, $\Sigma$, when the dimension of the covariance matrix is large relative to the sample size.

A number of approaches have been proposed for a stable estimation of the high-dimensional covariance matrix efficiently from a frequentist perspective. They are mainly based on developing regularized estimators through banding or tapering of the sample covariance matrix (Bickel and Levina, 2008; Wu and Pourahmadi, 2010). Alternative regularization methods have been developed by banding the cholesky or inverse cholesky factors (Wu and Pourahmadi, 2003; Huang et al., 2006; Levina et al., 2008); thresholding the sample covariance matrix (Bickel and Levina, 2008A; Cai and Liu, 2011) or regularizing the principal component analysis (Johnstone and Lu, 2009).  Leonard and Hsu (1992), Chiu et al. (1996), and Deng and Tsui (2010) modeled the matrix logarithm of the covariance matrix.  Ledoit and Wolf (2004) constructed a shrinkage estimator which is a linear combination of the sample covariance matrix and a pre-chosen matrix. Alternative parsimonious modeling methods by identifying zero off-diagonal elements in the covariance matrix or its inverse have been proposed by Yuan and Lin (2007), Friedman et al. (2008), Levina et al. (2008), Bien and Tibshirani (2011), among others. In a Bayesian framework, Wong et al. (2003) used a selection prior for off-diagonal elements of the partial correlation matrix to identify zeros in the inverse covariance matrix. Talluri et al. (2009) and Wang (2012) used the Bayesian graphical lasso priors for sparse inverse covariance matrix estimation.  Furthermore, the hyper-inverse Wishart (HIW) prior was employed for covariance selection given a decomposable Gaussian graphical model (Lauritzen, 1996; Giudici and Green, 1999; Armstrong et al., 2009), which was extended for nondecomposable graphical models by Giudici and Green (1999), Roverato (2002), Brooks et al. (2003),  and Atay-Kayis and Massam (2005).

Luo (2011) introduced a different covariance structure for high-dimensional datasets, which can be decomposed into the summation of a low-rank and sparse matrix as
\begin{align}
\Sigma = L + S, \label{eq:1}
\end{align}
where $L$ is a low rank symmetric component and $S$ is a sparse symmetric positive definite component. This decomposition of covariance matrices for dimension reduction has a wide range of applications including factor analysis, random effects and conditional covariance models, where the low rank component $L$ indicates that the variation of the random vector can mostly be explained by a small number of common factors or principal components, and the sparse part $S$ displays the variance/covariance between the variables conditional on these latent factors. The covariance matrices in the models are not sparse in themselves, and hence banding, thresholding, or other parsimonious modeling methods are not appropriate to apply directly to these models. However, parsimonious modeling could be used for the covariance of the residuals after removing latent factors. In contrast to the popular principal component or low rank approximation methods, this decomposition method has the ability to provide a full rank covariance estimator. Luo (2011) proposed a frequentist approach LOREC, which regularizes $\hat{\Sigma}$ by the Frobenius norm and uses a composite penalty on the trace norm of $L$ and the $l_1$ norm of $S$ to achieve the low-rank and sparse component estimation, respectively. This method provides a point estimate of the covariance matrix ignoring uncertainties from different sources. In this paper, we propose a  Bayesian approach to estimate the covariance matrix with the decomposition structure in equation (\ref{eq:1}). We also extend the method for graphical factor analysis.

We represent the $q \times q$ low-rank matrix $L$ utilizing a singular value decomposition (SVD) as follows
\begin{align}
L = M D_\tau M^T, \label{eq:2}
\end{align}
where $M \in \mathbf{R}^{q\times r^*}$, the diagonal matrix $D_\tau = \mathrm{diag}(\tau^2_1,...,\tau^2_{r^*}) \in \mathbf{R}^{r^* \times r^*}$ consists of singular values of $L$, and $r^*$ denotes the true rank of $L$. This decomposition ensures positive definiteness of $\Sigma$ and determines the rank of $L$, which is given by the dimension of $D_\tau$. When $L$ is low-rank, $r^* <<q$. This representation of the covariance matrix has the same structure as in a factor analytic model, where $M$ could be viewed as the latent factor loadings matrix, the singular values $\tau^2_k$'s as the variances of the latent factors.  In our Bayesian method, we jointly estimate $L$ and $S$ through the factor analytic model, which has an unknown number of factors and a sparse covariance matrix of the residuals. To estimate the rank of $L$, i.e. the number of true factors,  we introduce a binary indicator for each factor separating the factor selection and the singular value estimation, and we assign conjugate type priors such that most of the parameters have closed-form conditional posterior distributions, which facilitates the Markov chain Monte Carlo (MCMC) computation and also allows an automatic choice of the regularization parameters.

In statistical applications such as gene expression and financial data analysis, we often are interested in the graphical model of the variables, which is flagged by the zero pattern in the off-diagonal of $S$ inverse. In these cases, the sparsity in $C=S^{-1}$ instead of $S$ is desired for graphical model inference. For this purpose, we extend our method to a graphical factor model so that it achieves sparsity in estimating $C$. We employ a conjugate HIW prior on $S$ in the graphical factor model when $S$ is restricted to a decomposable graph. We further extend the model for unrestricted graphs by using a Bayesian graphical lasso selection prior. We show through simulations that the extended model can recover both the number of latent factors and the graphical model of the residuals successfully when the sample size is sufficient relative to the dimension.

The rest of the paper is organized as follows. In Section \ref{sec:model}, we describe our proposed Bayesian model for low rank and sparse covariance matrix estimation. In Section \ref{sec:simu1} we report results from simulation studies to assess the operating characteristics of our method. A real data analysis of gene expression data is included in Section \ref{sec:realdata}. In Section \ref{sec:extendmodel} we extend our method to develop a graphical factor model to handle data where selection of the latent factors as well as inference of the graphical model among variables are both of interest, and show the application of the graphical factor model to a gene expression dataset. We provide a discussion and conclusion in Section \ref{sec:diss}.

\section{Bayesian Low Rank and Sparse Covariance Model} \label{sec:model}

\subsection{Likelihood Model}

Consider a $q \times n$ data matrix $\mathbf{y}$, with each column vector $\mathbf{y}_i$ for $i=1,\ldots,n$ following an identical and independent (iid) Gaussian distribution
\begin{align}
\mathbf{y}_i \sim \mathcal{N}_q(\mathbf{0},\Sigma).       \label{eq:L1}
\end{align}

We assume that the covariance matrix $\Sigma$ could be decomposed as a sum of a low rank component $L$ and a sparse component $S$, with $L$ to be represented in the form of SVD as in formula (\ref{eq:2}). Hence we have
\begin{align}
\Sigma = M D_\tau M^T + S.  \label{eq:3}
\end{align}
Note that the representation of the covariance in formula (\ref{eq:3}) can be viewed from the standpoint of a latent factor analytic model:
\begin{align}
\mathbf{y}_i = M \mathbf{f}_i + \boldsymbol{\epsilon}_i,  \label{eq:5}
\end{align}
where $\mathbf{f}_i=(f_{1i}, \ldots, f_{r^*i})^T$ is the value of an $r^*$-dimensional random vector of latent factors in the $i^{\rm th}$ replicate, $M$ is the $q \times r^*$ latent factor loadings matrix, the diagonal elements in $D_\tau$ are the variances of $\mathbf{f}_i$, and $S$ is the covariance matrix of the $\boldsymbol{\epsilon}_i$.

Factor analytic models have been extensively studied for summarizing the variance and covariance patterns in multivariate data. With advances in computational tools, Bayesian methods for factor analysis have been rapidly developed. See, for example, Geweke and Zhou (1996), Aguilar and West (2000), and Rowe (2003) among others. Lopes and West (2004) explored the inference on the number of latent factors in a factor model with a reversible jump MCMC method. Other recent Bayesian factor analysis methods incorporated different modeling structures through the columns of the factor loadings matrix (Lopes and Carvalho, 2007; Carvalho et al., 2008). Hoff (2007) proposed a model based singular value decomposition method introducing the unknown rank scenario. Using the above decomposition, Fan et al. (2008), Hoff and Niu (2012) proposed a regression type estimator assuming $\mathbf{f}_i$ is observable. However, most of the methods assume that the covariance matrix of the residuals $S$ is diagonal. That is, all the association among the observed variables are exclusively contributed to the latent factors.

In our proposed method we assume that $S$ is a sparse covariance matrix. That is, we allow $S$ to be nondiagonal so that the variables could be correlated conditional on the latent factors. Grzebyk, Wild and Chouani\`{e}re (2004) gave a sufficient condition for the identification of a multi-factor model with correlated residuals as in (\ref{eq:3}). Since we do not know the rank of $L$, i.e. the number of latent factors $r^*$ in the model, we need to  estimate the number of factors and the variances of the factors jointly. To separate these two tasks, we introduce an extra binary indicator matrix $Z$. Our proposed Bayesian model is
\begin{align}
\Sigma = M (Z D_\tau) M^T + S,    \label{eq:4}
\end{align}
where $D_\tau$ is a diagonal matrix with positive diagonal elements $\tau^2_k$ for $k=1,\ldots,r$ for some $r > r^*$, and $Z$ is a diagonal matrix with binary entries $z_k \in \{0,1\}$ for $k=1,\ldots,r$ along the diagonal. While $\tau^2_k$ gives the variance of the $k^{th}$ latent factor, the indicator $z_k$ determines if the latent factor is included in the model. In this way, we separate the recovery of the rank and the estimation of the singular values of $L$. The rank of $L$, $r^*$, is only determined by the number of $1$'s in the diagonal of $Z$. Now the estimation of $r^*$ is equivalent to selecting the true number of latent factors in a factor analytic model. In our method, we choose a relatively large integer $r \leq q$ which is supposed to be much larger than $r^*$, and expect that the diagonal entries of $Z$ are sparse. If the estimates of $Z$ diagonals are not sparse, we increase the value of $r$.

We can rewrite the likelihood model in (\ref{eq:L1}) and (\ref{eq:4}) as the regression-type representation of a latent factor analysis model:
\begin{align}
\mathbf{y}_i & =M Z \mathbf{f}_i+\boldsymbol{\epsilon}_i, \text{ for }i=1,\ldots,n     \label{eq:L2} \\
f_{ki} & \sim \mathcal{N}(0,\tau_k^2),  \text{ for } k=1,\ldots,r \nonumber  \\
\epsilon_i & \sim \mathcal{N}_q(0,S),           \nonumber
\end{align}
If $z_k=1$, the $k^{th}$ factor, $k=1,\ldots,r$, is a true factor of the variables $\mathbf{Y}$ with variance $\tau^2_k$; otherwise, the $k^{th}$ factor is not included in the factor model. By rewriting the model in the form of a linear regression problem, we can assign conjugate priors to $M$, $Z$ and $D_\tau$, which leads to closed-form full conditional distributions of the parameters and facilitates the posterior sampling using a Gibbs algorithm.

\subsection{Prior Specification}
\subsubsection{Low Rank Component} \label{subsec:model11}

To complete model specification, we need to assign priors to the set of parameters $\{M,Z,D_\tau,S\}$ in the hierarchical likelihood model (\ref{eq:L2}), where $M,Z,D_\tau$ give the low rank component $L$, and $S$ is the sparse component. Let $m_{jk}$ be the element of $M$ in the $j^{th}$ row and $k^{th}$ column, and $\mathbf{M}_k=(m_{1k},\ldots,m_{qk})^T $ be the $k^{\rm th}$ column vector of $M$, which could be viewed as the loading vector of the factor $k$ on the variables. We assume that $M_k$ has a Gaussian prior:
\begin{align*}
\mathbf{M}_k \sim \mathcal{N}_q \left(\mathbf{0},\frac{1}{q} I_q \right), \quad  k = 1,\ldots,r,
\end{align*}
where $I_q$ is a $q \times q$ identity matrix. Note that for large values of $q$, the columns of $M$ are approximately orthogonal. Furthermore, we notice that $M$ and $D$ are unidentifiable. By assigning the prior variance of $m_{ik}$ to be $1/q$, we reduce the variability of $m_{ik}$, shift the variability to the single element $\tau^2_k$ in $D_\tau$, and obtain a relatively stable estimate of $L$.

The binary diagonal matrix $Z$ is modeled as
\begin{align*}
z_k \sim \mathrm{Bernoulli}(p_k),  \quad k = 1,\ldots,r,
\end{align*}
where $p_k$ is the prior probability of $z_k=1$. The values of $p_k$ determine the strength of the penalization that is assigned to the rank of $L$ as $r^*=rank(L)$ is equivalent to the number of $z_k$'s that equal $1$. Since $L$ is assumed to be low-rank, most of the prior probabilities are expected to be small or zero. We model these probabilities with the following hyper-prior distribution:
\begin{align*}
p_k & \sim (1-\pi) \mathcal{I}\{p_k=0\} + \pi \mathrm{Beta}(a_p,b_p),
\end{align*}
where $\mathcal{I}(\cdot)$ is the indicator function. The hyper-prior of $p_k$ is a Beta distribution mixed with a point mass at $0$ with probability $\pi$, where $\pi$ is drawn from a Beta prior distribution
$$\pi \sim \mathrm{Beta}(a_{\pi},b_{\pi}).$$
The sparseness of the diagonals of $Z$ is explicitly imposed through the hyperparameters $(a_p,b_p)$ and $(a_{\pi}, b_{\pi})$. When $a_{\pi}/(a_{\pi}+b_{\pi})\ll 1$, $p_k$ has a high prior probability to be zero; when $a_p/(a_p+b_p) \ll 1$, $p_k$ is still likely to be close to zero, if it is not exactly zero. To impose high penalization on the rank of $L$, we choose $(a_{\pi}, b_{\pi})=(1/q,1-1/q)$ and $(a_{\pi}, b_{\pi})=(1,r)$.

Each diagonal entry $\tau^2_k$ in $D_\tau$ corresponds to the variance of the $k^{th}$ factor. We can assign a conjugate Inverse-Gamma prior
\begin{align*}
\tau^2_k \sim \mathrm{IG}(a_\tau,b_\tau), \quad k = 1,\ldots,r,
\end{align*}
which leads to a closed form of the posterior conditional distribution.

\subsubsection{Sparse Component} \label{subsec:model12}

In order to achieve adaptive shrinkage of the sparse component $S$, we use a Bayesian  lasso prior for the sparse $S$ estimation. In the graphical lasso method (Yuan and Lin 2007; Friedman et al. 2008), an $l_1$ penalty term is assigned to $S$, which, in a Bayesian framework, is equivalent to a prior distribution as follows
\begin{align*}
S | \lambda \propto \prod_{j<j'} \exp\{-\lambda|S_{jj'}|\} \prod^q_{j=1} \exp\{-\frac{\lambda}{2}|S_{jj}|\} \mathcal{I}\{ S \in \mathcal{S}^+\},
\end{align*}
where $\mathcal{S}^+$ is the space of all positive definite $S$, and $\lambda>0$ is the regularization parameter. The prior is a joint density of exponential distributions on the diagonal elements $S_{jj}$, $j=1,\ldots,q$, and Laplace distributions on the off-diagonal elements $S_{jj'}$, $j<j'$, with the constraint $S \in \mathcal{S}^+$. However, the Bayesian lasso method shrinks but does not set the off-diagonal elements to exact zeros, which is desired for the sparse $S$ estimation. To this end, we modify the method by placing a point mass at $0$ in the Laplace prior for the off-diagonal elements. The prior on $S$ is then as follows:
\begin{align*}
S | \lambda, \boldsymbol\rho  \sim & con_{\lambda,\rho}^{-1} \prod_{j<j'} \Big[ (1-\rho_{jj'}) \mathcal{I}\{S_{jj'}=0\} + \rho_{jj'} \mathrm{Laplace}(\lambda) \Big] \cdot \prod^q_{j=1} \mathrm{Exp}(\frac{\lambda}{2}) \cdot   \mathcal{I}\{ S \in \mathcal{S}^+\},
\end{align*}
where $\mathrm{Laplace}(\lambda)$ is a Laplace distribution with $\lambda$ to be the rate parameter, and $ \mathrm{Exp}(\frac{\lambda}{2})$ is a exponential distribution with rate $\frac{\lambda}{2}$. Here, $con_{\lambda,\rho} = \int_{S \in \mathcal{S}^+}  \prod_{j<j'} \Big[ (1-\rho_{jj'}) \mathcal{I}\{S_{jj'}=0\} + \rho_{jj'} \mathrm{Laplace}(\lambda) \Big] \prod^q_{j=1} \mathrm{Exp}(\frac{\lambda}{2}) dS$ is the normalizing constant, which depends on $\lambda$ and $\boldsymbol\rho = \{\rho_{jj'},j<j'\}$, and is analytically intractable.

In this construction, the hyper-parameter $\lambda$ shrinks the covariance elements toward zero, while $\rho_{jj'}$ controls the probability that the $(j,j')$ element will be enforced to be a zero. In the Bayesian framework, we can assign hyperpriors to $\lambda$ and $\rho_{jj'}$ and implement the regularization parameters in the MCMC algorithm for posterior inference. However, since the normalizing term $con_{\lambda,\rho}$ cannot be evaluated analytically, standard prior distributions on the hyper-parameters will lead to intractable posterior distributions. Following Wang (2012), we consider an extension of the conjugate priors on $\lambda$ and $\boldsymbol\rho$
\begin{align*}
(\lambda, \boldsymbol\rho) & \sim  con_{\lambda,\rho} \mathrm{Gamma}(a_\lambda, b_\lambda) \prod_{j<j'} \mathrm{Beta}(a_\rho, b_\rho).
\end{align*}
The prior distribution is proper as $con_{\lambda,\rho} < \int_{S \in \mathbb{R}^{q(q+1)/2}}  \prod_{j<j'} \Big[ (1-\rho_{jj'}) \mathcal{I}\{S_{jj'}=0\} + \rho_{jj'} \mathrm{Laplace}(\lambda) \Big] \\ \prod^q_{j=1} \mathrm{Exp}(\frac{\lambda}{2}) dS = 1$. Given the joint hyperprior, the full conditional distributions for $\lambda$ and $\rho_{jj'}$ are independent Gamma and Beta distributions, respectively.
In our experiments, we specify $(a_\lambda,b_\lambda)=(1,1)$ for a diffuse prior for $\lambda$, and $(a_\rho,b_\rho)=(0.5,0.5)$ for a noninformative prior of $\rho_{jj'}$.

The complete hierarchical model can be summarized as
\begin{align}
& \left. \quad
\begin{aligned}
   \mathbf{y}_i & \sim \mathcal{N}_q(M Z \mathbf{f}_i,S),            \quad  i=1,\ldots,n    \quad                                  \\
         f_{ki} & \sim \mathcal{N}(0,\tau_k^2),             \quad  k=1,\ldots,r
\end{aligned}
\right\}        \label{eq:likelihood}\\
& \left. \quad
\begin{aligned}
   \mathbf{M}_k & \sim \frac{1}{q} \mathcal{N}_q(\mathbf{0},\frac{1}{q}I_q),                         \\
          z_{k} & \sim \mathrm{Bernoulli}(p_k),                                            \\
            p_k & \sim (1-\pi) \mathcal{I}\{p_k=0\} + \pi \mathrm{Beta}(a_p,b_p),   \quad   \\
            \pi & \sim \mathrm{Beta}(a_{\pi},b_{\pi}),                                       \\
       \tau^2_k & \sim \mathrm{IG}(a_\tau,b_\tau),
\end{aligned}
\right\}        \label{eq:L_hierarchy}\\
& \left.
\begin{aligned}
        S | \lambda, \boldsymbol\rho  & \sim  con_{\lambda,\rho}^{-1} \prod_{j<j'} \Big[ (1-\rho_{jj'}) \mathcal{I}\{S_{jj'}=0\} + \rho_{jj'} \mathrm{Laplace}(\lambda) \Big] \\
						& \quad \cdot \prod^q_{j=1} \mathrm{Exp}(\frac{\lambda}{2}) \cdot   \mathcal{I}\{ S \in \mathcal{S}^+\},       \\
       (\lambda, \boldsymbol\rho) & \sim  con_{\lambda,\rho} \mathrm{Gamma}(a_\lambda, b_\lambda) \prod_{j<j'} \mathrm{Beta}(a_\rho, b_\rho),
\end{aligned}
\right\}        \label{eq:S_hierarchy1}
\end{align}
where $j$ denotes the variable, $i$ denotes the sample, $k$ denotes the latent factor, $j=1,\ldots,q$, $i=1,\ldots,n$, and $k=1,\ldots,r$.

\subsection{Posterior Inference}  \label{subsec:model1_postsample}

In this section, we present the full conditional posterior distributions and a framework to carry out MCMC calculations. Note that the full conditional distributions of most parameters are available in a closed form, allowing for a straightforward Gibbs sampling algorithm, except for the off-diagonal elements in the sparse matrix $S$.  In this case, we employ an independent Metropolis Hastings (MH) sampler within the Gibbs sampler to generate posterior samples of  the off-diagonal elements in $S$.


\begin{enumerate}
\item {\bf Sampling the factor loadings matrix $M$:}

Let $\mathbf{M}_k=(m_{1k},\ldots,m_{qk})^T$ be the $k^{th}$ column vector of $M$, and $M_{(-k)}$ be the matrix of $M$ excluding the $k^{th}$ column. The full conditional distribution of $\mathbf{M}_k$ for $k=1,\ldots,r$ is
\begin{align*}
\mathbf{M}_k|y,M_{(-k)},f,S & \sim  \mathcal{N}_q(\boldsymbol{\mu}^M_k,\Sigma^M_k),
\end{align*}
where $\Sigma^M_k = \Big\{ S^{-1}( \sum^n_{i=1}f^2_{ki} )+ qI_q \Big\}^{-1}$, and $\boldsymbol{\mu}^M_k = \Sigma^M_k S^{-1}(y-M_{ (-k)}f_{(-k)\cdot})\mathbf{f}_{k\cdot}^T$. Hence we can draw samples of each column of $M$ from a multivariate Gaussian distribution.

\item {\bf Sampling the random factors $f$:}

Let $f$ be the $r \times n$ matrix with $f_{ki}$ to be the value of $k^{th}$ factor in $i^{th}$ replicate. Then $\mathbf{f}_{k\cdot} = (f_{k1},\ldots,,f_{kn})$ is the $k^{th}$ row vector of $f$, and  $f_{(-k)\cdot}$ denotes the matrix of $f$ excluding the $k^{th}$ row. Note $\mathbf{f}_{k\cdot}$ could be viewed as the unobserved values of the random factor $k$ in the $n$ replicates. The full conditional distribution of the transpose of $\mathbf{f}_{k\cdot}$ for $k=1,\ldots,r$ is
\begin{align*}
\mathbf{f}^T_{k\cdot} | y,M,f_{(-k)\cdot},S,z_k,\tau^2_k & \sim  (1-z_k)\mathcal{N}_q(0,\tau_k^2 I_n) +
                            z_k \mathcal{N}_n(\boldsymbol{\mu}^f_k, \sigma^f_k I_n),
\end{align*}
where $\sigma^f_k = \Big( M^T_k S^{-1} M_k +\tau^{-2}_k \Big) ^{-1}$, and $\boldsymbol{\mu}^{f}_k = \sigma^{f}_k \Big\{ M^T_kS^{-1}(y-M_{(-k)}f_{(k)\cdot}) \Big\}^T$. Hence we can draw samples of each row of $f$ from a mixture of two multivariate Gaussian distributions.

\item {\bf Sampling the binary diagonal matrix $Z$:}

The full conditional of each diagonal element of $Z$, $z_k$, is a Bernoulli distribution:
\begin{align*}
z_k | y,M,f,S,\tau^2_k,p_k \sim & Bernoulli(p^*_k),
\end{align*}
where $p^*_k =  1- (1-p_k)/\Big\{ 1-p_k+p_k\cdot \Big( \frac{\sigma^f_k}{\tau^2_k} \Big)^{\frac{n}{2}} \cdot \exp \Big( -\frac{1}{2} \frac{(\boldsymbol{\mu}^f_k)^T \boldsymbol{\mu}^f_k}{\sigma^f_k} \Big) \Big\}$.

\item {\bf Sampling the probabilities $p_k$ and $\pi$:}

The full conditional of $p_k$ for $k=1,\ldots,r$ is
\begin{align*}
p_k | z_k=1 & \sim  Beta ( a_p+1, b_p ), \\
p_k | z_k=0 & \sim  (1-\pi^*) \mathcal{I}\{p_k=0\} + \pi^*Beta(a_p, b_p+1),
\end{align*}
where $\pi^* =  \frac{\pi b_p}{a_p+b_p-\pi a_p}$.

The full conditional of $\pi$ is
\begin{align*}
\pi \sim Beta \left( a_{\pi}+\sum_k \mathcal{I}\{p_k=0\}, b_{\pi}+ \sum_k \mathcal{I}\{p_k \neq 0 \} \right).
\end{align*}

\item {\bf Sampling the positive diagonal matrix $D_\tau$:}

The full conditional of each diagonal element of $D_\tau$, $\tau^2_k$, for $k=1,\ldots,r$ is
\begin{align*}
\tau^2_k | z_k, \mathbf{f}_{k \cdot} \sim  (1-z_k) \mathrm{IG}(a_\tau,b_\tau) + z_k \mathrm{IG} \left( a_\tau+\frac{n}{2}, b_\tau+\frac{\mathbf{f}_{k \cdot} \mathbf{f}^T_ {k\cdot}}{2} \right).
\end{align*}

\item {\bf Sampling the diagonal elements $S_{jj}$:}

The full posterior conditional density of the diagonal element $S_{jj}$, for $j=1,\ldots,q$, is
\begin{align*}
p(S_{jj} | \cdot) \propto \det(S)^{-\frac{n}{2}} \exp \left\{-\frac{tr(\Lambda S^{-1})}{2}-\frac{\lambda S_{jj} }{2} \right\} \mathcal{I} \{S \in \mathcal{S}^+ \},
\end{align*}
where $\Lambda=(y-Mf)(y-Mf)^T$. Without loss of generality, suppose that $j=q$. By the properties of matrix inverse and matrix determinant, we have
\begin{align*}
p(S_{jj} | \cdot) \propto (S_{jj}-c)^{-\frac{n}{2}} \exp \{-\frac{d}{2}(S_{jj}-c)^{-1} -\frac{\lambda}{2} S_{jj} \} \mathcal{I} \{S_{jj}>c\},
\end{align*}
where {\small $c = S_{j,-j} S^{-1}_{-j,-j} S_{-j,j}$, and $d = [\begin{array}{lr} S_{j,-j}S^{-1}_{-j,-j} & -1 \end{array} ] \cdot \Lambda \cdot [\begin{array}{cc} S_{j,-j}S^{-1}_{-j,-j} & -1 \end{array} ]^T$}. Note that the indicator function $\mathcal{I} \{S_{jj}>c\}$ ensures the positive definiteness of $S$ conditional on the other elements in $S$. The derivation of the full conditional distributions for the elements of $S$ is detailed in Appendix. The above distribution is in a closed form. We now transform $S_{jj}$ to $\nu=S_{jj}-c$, then the conditional density of $\nu$ is
\begin{align*}
p(\nu|\cdot) \propto \nu^{-\frac{n}{2}} \exp\{-(d/\nu+\lambda \nu)/2\} \mathcal{I} \{\nu>0\},
\end{align*}
which is a generalized inverse Gaussian (GIG) distribution with parameters $(1-n/2,d,\lambda)$. Therefore, at each MCMC iteration, we can draw a sample of $\nu$ from the GIG distribution, and obtain $S_{jj}=\nu+c$.

\item {\bf Sampling the off-diagonal elements $S_{jj'}$: }

The full posterior conditional density of the off-diagonal element $S_{jj'}$, $j<j'$ is
{\footnotesize
\begin{align*}
p(S_{jj'} | \cdot) \propto \det(S)^{-\frac{n}{2}} \exp \Big\{-\frac{tr(\Lambda S^{-1})}{2} \Big\} \Big[(1-\rho_{jj'}) \mathcal{I}\{S_{jj'}=0\} + \rho_{jj'} \frac{\lambda}{2} \exp(-\lambda |S_{jj'}|) \Big]  \mathcal{I} \{S \in \mathcal{S}^+ \} .
\end{align*}
}
Without loss of generality, suppose that $j=q-1$ and $j'=q$. By the properties of matrix inverse and matrix determinant, we have
{\footnotesize
\begin{align*}
p(S_{jj'}|\cdot) \propto & \left\{ 1-\frac{(S_{jj'}-B_{12})^2}{(S_{jj}-B_{11})(S_{j'j'}-B_{22})}\right\}^{-\frac{n}{2}} \\
               & \cdot     \exp \left\{ -\frac{1}{2}\frac{(S_{j'j'}-B_{22})D_{11}+(S_{jj}-B_{11})D_{22}-2(S_{jj'}-B_{12})D_{12})}
                          {(S_{jj}-B_{11})(S_{j'j'}-B_{22})-(S_{jj'}-B_{12})^2}  \right\} \\
              & \cdot \Big[ (1-\rho_{jj'}) \mathcal{I} \{S_{jj'}=0\} + \rho_{jj'} \frac{\lambda}{2} \exp(-\lambda |S_{jj'}|) \Big]
			\cdot \mathcal{I} \left\{(S_{jj'}-B_{12})^2<(S_{jj}-B_{11})(S_{j'j'}-B_{22}) \right\} ,
\end{align*}
}
where {\footnotesize $B=S_{jj',-(jj')}S^{-1}_{-(jj')}S_{-(jj'),jj'}$}, and {\footnotesize $D = [S_{jj',-(jj')}S^{-1}_{-(jj')},\text{ } -I_2] \cdot \Lambda \cdot  [S_{jj',-(jj')}S^{-1}_{-(jj')},\text{ } -I_2]^T$}. We transform $S_{jj'}$ to $\nu=S_{jj'}-B_{12}$ and let $a=S_{jj}-B_{11}$, $b=S_{j'j'}-B_{22}$, then the conditional density of $\nu$ is
\begin{align*}
p(\nu|\cdot) \propto & (1-\rho_{jj'}) g(-B_{12}) \mathcal{I} \{\nu=-B_{12}, \nu^2<ab\}  + \frac{\rho_{jj'} \lambda}{2} g(\nu),
\end{align*}
where $g(\nu) =  (1-\frac{\nu^2}{ab})^{-n/2} \exp \left\{ -\frac{bD_{11}+aD_{22}-2D_{12}\nu}{2(ab-\nu^2)} -\lambda |\nu+B_{12}| \right\} \mathcal{I} \{\nu^2<ab\}$. Note that the positive definite constraint on $S$ is ensured by the indicator $\mathcal{I} \{\nu^2<ab\}$. That is, the posterior distribution of $\nu$ is only over the region $(-\sqrt{ab},\sqrt{ab})$.

The continuous part of the conditional distribution of $\nu$, $g(\nu)$, cannot be sampled directly. Furthermore, $g(\nu)$ is nonconcave and therefore the sampler may be trapped locally if we use the random-walk MH algorithm within the Gibbs sampling. Since $g(\nu)$ only has density over $(-\sqrt{ab},\sqrt{ab})$ and is zero elsewhere, we construct a piecewise uniform proposal distribution approximating $g(\nu)$.

We choose $\kappa-1$ equally spaced grids between ($-\sqrt{ab}$, $\sqrt{ab})$, $-\sqrt{ab}=\nu_0<\nu_1<\cdots<\nu_{\kappa}=\sqrt{ab}$, which divide the domain of $\nu$ into $\kappa$ intervals of width $2\sqrt{ab}/\kappa$. The piecewise uniform is as follows:
\begin{align*}
g_a(\nu) = \left\{ \begin{array}{rl}
            g(\frac{\nu_0 + \nu_1}{2}) & \mbox{ if $\nu_0 < \nu \leq \nu_1$ } \\
            g(\frac{\nu_1 + \nu_2}{2}) & \mbox{ if $\nu_1 < \nu \leq \nu_2$ } \\
            \cdots & \\
            g(\frac{\nu_{\kappa-1} + \nu_\kappa}{2}) & \mbox{ if $\nu_{\kappa-1} < \nu < \nu_\kappa$ }
            \end{array} \right.
\end{align*}

The independent MH proposal for generating $\nu=S_{jj'}-B_{12}$ is given by
\begin{align*}
q(\nu|\cdot) \propto & (1-\rho_{jj'}) g(-B_{12}) \mathcal{I} \{\nu=-B_{12}, \nu^2<ab \}  + \frac{\rho_{jj'} \lambda}{2} g_a(\nu).
\end{align*}

Samples of $\nu$ could be generated using an inverse-CDF method from the proposal distribution and the proposal $\nu^*$ is accepted with the probability
\begin{align*}
\alpha = \min \left\{ 1, \frac{p(\nu^*)}{q(\nu^*)} \Big/ \frac{p(\nu^c)}{q(\nu^c)} \right\},
\end{align*}
where $\nu^c = S^c_{jj'}-B_{12}$ denotes the current state of $\nu$. The piecewise uniform proposal distribution avoids the local-trap problem and can be sampled easily using an inverse-CDF method. Furthermore, $q(\nu | \cdot)$ approximates the distribution $p(\nu|\cdot)$ more accurately with the increases of the number of grids. Based on our simulations, $100$ grids are enough for a fast convergence of $S_{jj'}$.

\item {\bf Sampling the shrinkage parameter $\lambda$:}

The full conditional of $\lambda$ is
\begin{align*}
\lambda|S \sim \mathrm{Gamma}(a_\lambda+m, b_\lambda+\sum_{j<j'}|S_{jj'}|+\frac{1}{2}\sum_{j=1}^q S_{jj}),
\end{align*}
where $m$ equals to size of the set $\{(j,j'): j \leq j', S_{jj'} \neq 0 \}$.

\item {\bf Sampling the selection parameters $\rho_{jj'}$:}

The full conditional of $\rho_{jj'}$ for $j<j'$ is
\begin{align*}
\rho_{jj'} | S \sim \mathrm{Beta}(a_\rho+\mathcal{I} \{S_{jj'} \neq 0\}, b_\rho+\mathcal{I} \{ S_{jj'}=0 \}).
\end{align*}

\end{enumerate}

\section{Simulation Studies}  \label{sec:simu1}

We conducted a detailed simulation study to evaluate the operating characteristics of our method. We considered three covariance models to generate the data:

\begin{itemize}
\item Model 1: $\Sigma = MDM^T+I$, where $M \in \mathbf{R}^{q \times 3}$ with orthonormal columns, and $D=\mathrm{diag}(8,8,8) \cdot (q/n)$. This covariance model comes from a factor model with independent residuals.

\item Model 2: $\Sigma = 0.3\mathbf{11}^T+S$, where  $S$ is block diagonal with each square block matrix $B$ of dimension 5, and $B=0.7\mathbf{11}^T+0.3I$. This covariance matrix simulates a random effect model, with the covariance of the residuals to be block diagonal.

\item Model 3: $\Sigma = MDM^T+S$, where $M$ and $D$ are the same as in model 1, and $S$ is a block diagonal matrix as in model 2. This covariance model comes from a factor model with a block diagonal covariance matrix of the residuals.

\end{itemize}

For each model, 50 observations were generated from the multivariate Gaussian distribution $\mathcal{N}_q(\mathbf{0},\Sigma)$ with varying dimensions $q=50$, $100$, and $200$. We compared our proposed Bayesian model for low-rank and sparse covariance decomposition with the frequentist LOREC method (Luo, 2011) in estimating the covariance matrices as well as recovering the rank of $L$ and sparsity of $S$. The estimates of the parameters using the Bayesian method were based on the posterior samples of 10,000 iterations after 5000 burn-in iterations. The number of latent factors was estimated by the mode of its posterior distribution, and the support of the sparse $S$ was selected by a false discovery rate (FDR)-based model selection algorithm (Bonato et al. 2010) with an FDR of 0.20. The tuning parameters for the LOREC estimators are picked by 5-fold cross validation using the Bregman divergence loss as in Luo (2011).

Table~\ref{table2} compares the performance of covariance estimation using our Bayesian method, Luo's LOREC method, and the sample covariance over 20 replicates measured by the $l_1$ norm and the Frobenius norm. The two matrix norms are defined as follows: Let $X=(X_{ij})$ be any matrix; $|X|_1=\sum_i \sum_j|X_{ij}|$ gives the $l_1$ norm, and $|X|_F = \sqrt{\sum_i \sum_j X^2_{ij}}$ gives the Frobenius norm. The Bayesian method overall is comparable to the LOREC estimator and is better than the sample covariance estimator. While the LOREC estimator performs better for the random effect model, our Bayesian estimator has lower loss for the factor model with independent residuals. Both of them are better than the sample covariance in all models.


Table~\ref{table1} summarizes the inference results in the recovery of rank of $L$ and sparsity of $S$. The table shows that the Bayesian estimator can recover the true rank of the low rank components with high frequencies for all the three models, with the successful recovery rates close to the LOREC estimator. Furthermore, our method has much lower false positive rates in support recovery of $S$ when $S$ is non-diagonal, at the price of a little higher false negative rates. The above results indicate that our method can recover both the rank and the sparsity of the two components with high frequencies.


\section{Covariance Estimation For Gene Expression Dataset}  \label{sec:realdata}

In this section, we applied the Bayesian low rank and sparse decomposition model to estimating the covariance of a gene expression dataset from Stranger et al. (2005). The dataset is composed of 60 unrelated individuals of Northern and Western European ancestry from Utah (CEU) of which gene expression levels were measured throughout the genome. We considered 100 genes in the dataset which are most variable among all the genes available in the gene expression profile. Thus we had $n=60$ and $q=100$ in our dataset.

\begin{figure}
\centering
\centerline{\includegraphics[width=6in]{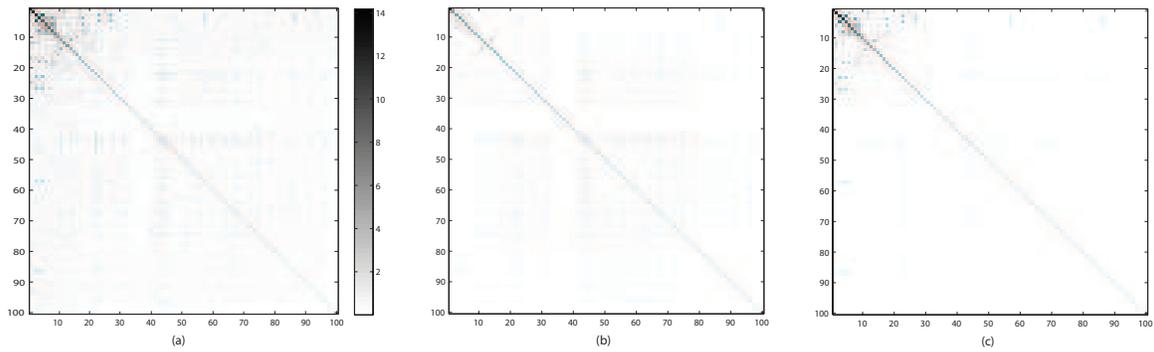}}
\caption{Heatmaps of the absolute value of covariance estimates by (a) sample covariance (b) Bayesian decomposition method (c) LOREC estimator.}
\label{figure1}
\end{figure}

We estimated the covariance matrix using our Bayesian method and compared it to the LOREC estimator and the sample covariance. Figure \ref{figure1} shows a heatmap showing the absolute intensities of three covariance matrix estimates. Compared to the sample covariance matrix, the LOREC estimator regulates the sample covariance estimate by shrinking all the off-diagonal elements uniformly, whereas the Bayesian decomposition model shrinks more of the elements with strong signals on the top left corner while keeps the abundant elements with moderate signals at the same time.

\begin{figure}
\centering
\centerline{\includegraphics[width=6in]{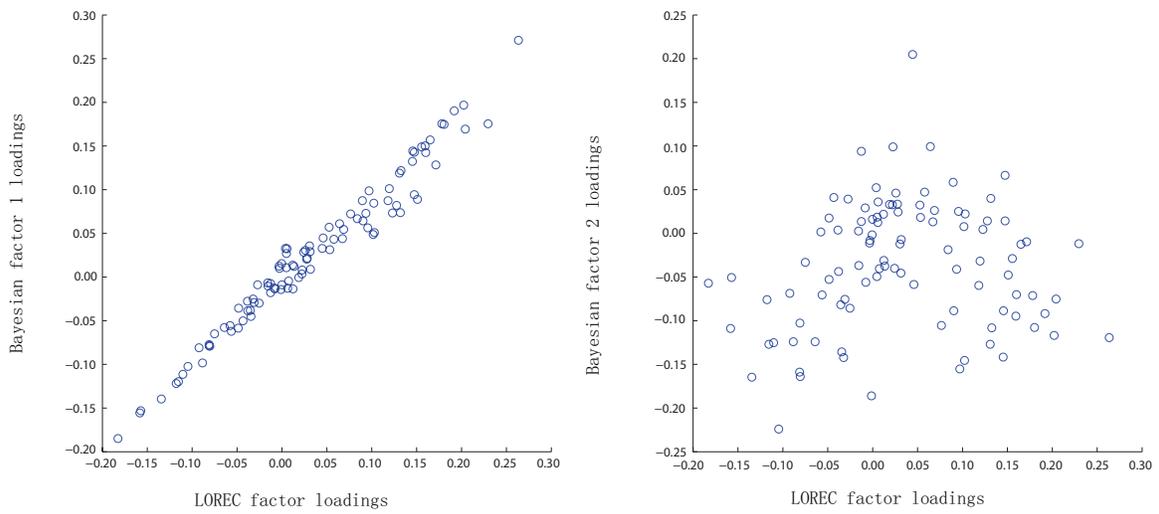}}
\caption{Scatter plots of the single factor loadings identified by LOREC versus the two factors loadings identified by the Bayesian decomposition model. The correlation between the two factor loadings vectors on the left is 0.98, and the correlation between the two factors on the right in 0.12.}\label{figure2}
\end{figure}

The Bayesian method and the LOREC estimator also decompose the covariance matrix into low rank and sparse components in the gene expression data. The LOREC estimator identifies a rank 1 component, and our Bayesian method identifies a low rank component of rank 2. The singular vector of a rank 1 component is equivalent (up to a multiplying constant) to the loading in a single factor model, and therefore we obtain the single loading vector from the rank 1 component of the LOREC estimator. We also obtain the loadings matrix corresponding to the two random factors identified by our Bayesian decomposition model. Figure~\ref{figure2} shows the scatter plots of the two loadings by the Bayesian model versus the single loading by the frequentist LOREC method, one of which corresponds to a correlation close to 1. It suggests that the Bayesian model identifies one latent factor with a similar loading as the LOREC estimator.

\begin{figure}
\centering
\centerline{\includegraphics[width=6in]{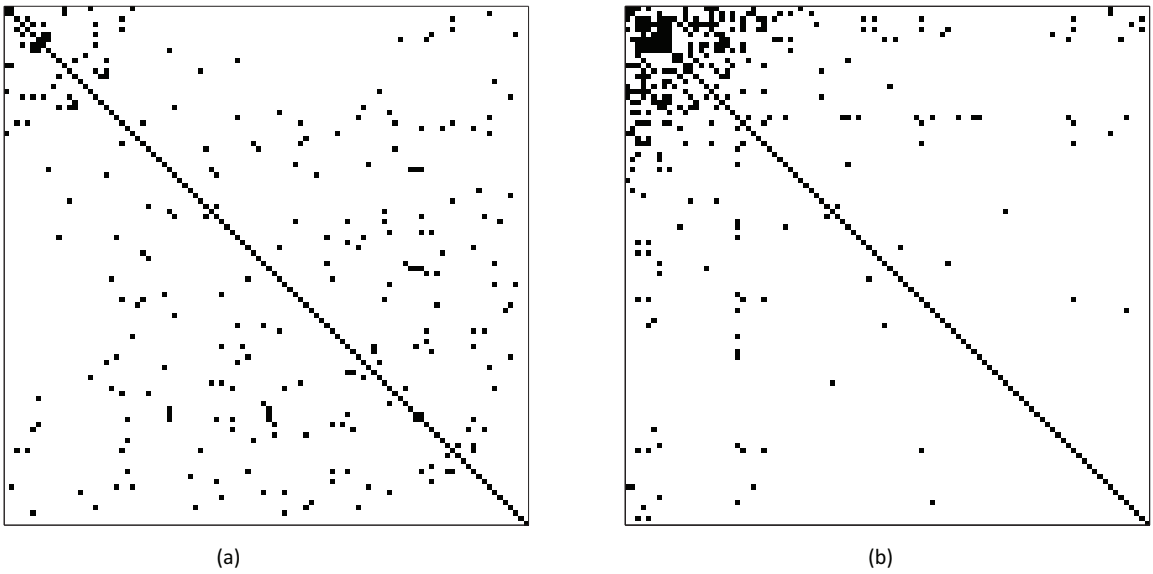}}
\caption{Matrix plot indicating the sparse support of the residual covariance component by (a) Bayesian decomposition method (b) LOREC estimator.}\label{figure3}
\end{figure}

Figure~\ref{figure3} displays the sparse support of the residual covariance component obtained by the Bayesian decomposition method as well as the LOREC estimator. The LOREC estimator identifies nonzero correlations predominantly on the left corner of the sparse component, whereas the Bayesian decomposition model detects nonzero correlations overspreading the sparse matrix. This difference in the support of the sparse component agrees with the patterns in the covariance estimators plotted in Figure~\ref{figure1}.

\section{Bayesian Graphical Latent Factor Model}   \label{sec:extendmodel}

In many applications such as gene expression data and financial data analysis, we are often interested in the inference of the graphical Markov model as they represent the conditional dependence associations among the set of observed variables. In these applications, it is generally accepted that the variation of a variable is directly regulated by a small subset of other variables, and many approaches including HIW and graphical lasso have been proposed for sparse estimation of the inverse covariance or concentration matrix as mentioned at the beginning of the paper. However, the associations among the observed variables could be, in many cases, partially contributed to some unobserved latent factors. For example, in gene expression data analysis, the genes included in the data might be commonly regulated by certain unknown genes or environmental factors lying upstream of the signaling pathway, whose levels cannot be measured. The existence of such latent factors could mask the true conditional independence structures among the observed variables. For this purpose, we are interested to identify the latent factors and infer the sparse graphical Markov model among the observed variables after removing the effects of these latent factors.

As shown in the  likelihood model (\ref{eq:L2}), we represented the covariance model as a latent factor model, where we assume that the covariance matrix $S$ of the residuals $\epsilon_i$  is sparse, and use a Bayesian  lasso selection prior to achieve sparsity in $S$. In this section, we extend the latent factor model given by (\ref{eq:L2}) by assuming that the concentration matrix of the residuals, $C=S^{-1}$, is sparse, whose nonzero pattern in the off-diagonal elements gives the conditional dependence arising out of a graphical model of the residuals. The extended latent factor model with sparse $C$ is a sparse graphical factor model with the number of factors unknown.

Giudici (2001) introduced the concept of a graphical factor model, which generalizes factor analytic models by allowing the concentration matrix of the residuals to have non-zero off-diagonal elements. He used an HIW prior (Dawid and Lauritzen 1993) for inference on the concentration matrices restricted to decomposable graphical models, and assigned a uniform prior on all decomposable graphs. We make the following contributions in our proposed model. Firstly, we recover the number of factors as well as the graphical model in a graphical factor model framework. Secondly, we propose a novel prior on the decomposable graphs for the HIW method, which induces adaptive sparsity in the inferred graphical model. Finally, we extend the method  for unrestricted graphs by using a Bayesian graphical lasso selection method. This framework allows for additional flexibility both in the aspect of the analysis of latent factors as well as for  graphical models.

\subsection{The Graphical Factor Model with Unknown Number of Factors}

Before introducing the graphical factor model, we first describe the notations in a graphical model. Let $Y$ be a $q-$dimensional vector of random variables. A conditional independence graph is a pair of $G=(V, E)$ with the vertex set $V=\{1,...,q\}$ and the edge set $E \subseteq V \times V$. Nodes $j$ and $j'$ are adjacent or connected in $G$ if $(j,j') \in E$, whereas $j$ and $j'$ are conditionally independent if $(j,j') \notin E$. A graph $G$ with $E=V \times V$ is called a complete graph. Complete subgraphs $P \subset V$ are called cliques; the joint subset of two cliques is called a separator denoted by $Q$. If a graph $G$ could be partitioned into a sequence of subgraphs $(P_1,Q_2,P_2,...,P_K)$ such that $V=\bigcup_k P_k$ and $Q_k=P_{k-1}\bigcap P_{k}$ are complete for all $k=1,...,K$, $G$ is called a decomposable graph (Lauritzen 1996). For a covariance matrix $S$ of the variables $Y$, let $C=S^{-1}$ be the concentration matrix. Nodes $j$ and $j'$ are conditionally independent, given other nodes, if and only if $C_{jj'}=0$. Thus, the graph $G$ is given by the configuration of nonzero off-diagonal elements of $C$: $E=\{(j,j'): C_{jj'} \neq 0 \}$.

In a classical factor model as in formula (\ref{eq:5}), there is a pre-specified number of factors $r^*$ and the residuals are assumed to be independent with the covariance matrix $S=\mathrm{diag}(s_{11},\ldots,s_{qq})$. We relax the two assumptions in our model as follows: (i) the number of underlying factors, $r^*$, is unknown, and is thought to be much smaller than some pre-specified integer $r$. Since the number of common factors is usually small, we pick a moderate to larger value of $r$ and expect only a small fraction of the factors are selected. (ii) The inverse covariance/concentration matrix of the residuals, $C=S^{-1}$, is a sparse matrix allowing to be nondiagonal. That is, we allow nonzero off-diagonal elements in the concentration matrix $C$ so that the unobserved variables could be dependent on each other conditional on the latent factors.

A sufficient condition for identification of a graphical factor model with a single factor and multiple factors is given in Stanghellini (1997) and Guidici (2001) respectively. However, from a Bayesian viewpoint, identification is of less theoretical concern but more important for posterior computation as discussed in Guidici (2001). When the graphical factor model is unidentifiable (with more than one solutions), the likelihood would be flat, and the posterior distribution of parameters would be multimodal except for extremely informative priors.

Our objective is to select the true number of factors out of $r$ candidate factors, and to recover the sparse graphical model of the residuals represented by the nonzero pattern in $C$ as well. For factor selection, as described in Section \ref{sec:model}, we introduce binary indicators Z to determine if each candidate factor is included or excluded from the factor model, and utilize the same modeling method as in the Bayesian covariance decomposition model for factor selection and loadings estimation in the graphical factor mode.  That is, we assign the same priors for $M$ ,$Z$, and $D_\tau$ as in formulas (\ref{eq:L_hierarchy}). Hence the graphical factor model has the same likelihood as the Bayesian covariance decomposition model in formulas (\ref{eq:likelihood}), and the same hierarchy for factor estimation as in formulas (\ref{eq:L_hierarchy}). The difference primarily lies in the assumption and the modeling method for $S$, the covariance matrix of the residuals. In the following section, we will mainly present the methods of modeling $S$, whose inverse matrix is assumed to be sparse.

\subsection{Bayesian Hierarchical Model for Decomposable Graphs} \label{subsec:model2}

In this section, we focus on the modeling method when the graphical models of the residuals are restricted to be decomposable. The Bayesian model for unrestricted graphs will be discussed in the next section.

When the graphical model of the residuals is restricted to be decomposable, we allow the covariance matrix of the residuals $S$ to follow a mixture of HIW priors over decomposable graphs as
\begin{align*}
S \sim & \mathrm{HIW}(G,\delta,\Phi), \\
G \sim & p(G),
\end{align*}
where $\delta \in \mathbf{R^+}$ is a fixed degree-of-freedom parameter,  $\Phi$ is a symmetric positive-definite scale matrix, and $p(G)$ is the mixing prior over decomposable graphs. The HIW distribution was introduced by Dawid and Lauritzen (1993) with the probability density function (pdf) given by
\begin{equation*}
p(S|G,\delta,\Phi)=\frac{\prod^K_{k=1}p(S_{P_k}|\delta,\Phi_{P_k})}{\prod^K_{k=2}p(S_{Q_k}|\delta,\Phi_{Q_k})},
\end{equation*}
where $P_k$ and $Q_{k}$ are the cliques and separators of the graph $G$ respectively. The term $p(S_{P_k}|\delta,\Phi_{Q_k})$ denotes the inverse Wishart (IW) density of $S_{P_k}\sim \mathrm{IW} (\delta, \Phi_{P_k})$ with the pdf
\begin{equation*}
p(S_{P_k}|\delta,\Phi_{P_k}) \propto |S_{P_k}|^{-(\delta/2+|P_k|)} \exp \Big\{-\frac{1}{2}tr(S_{P_k}^{-1}\Phi_{P_k}) \Big\}.
\end{equation*}
The HIW distribution is a conjugate prior distribution for the covariance matrix $S$. Specifically, if $q$-dimensional random vectors $\mathbf{x_1},\ldots,\mathbf{x_n}$ follow an iid multivariate normal distribution $\mathcal{N}_p(\mathbf{0}, S)$, and $S$ follows $\mathrm{HIW}(G,\delta,\Phi)$, the posterior of $S$ is $S |\mathbf{x},G \sim \mathrm{HIW}(G, \delta+n, \Phi+\mathbf{x}^T\mathbf{x})$. The closed form of the posterior distribution for $S$ plays a key part in the posterior inference based on an MCMC algorithm. In our model, we consider $\delta=3$ as reflecting the lack of prior information on $S$, and specify $\Phi=I_q$.

Let $e_{ij}$ be a binary indicator denoting whether the $(i,j)$ is included or excluded from the graphical model. One option for the mixing prior $p(G)$ is to assign an independent prior probability of an edge, $p(e_{ij})$, to each pair of nodes $(i,j)$, so that $\pi(G)=\prod_{(i,j) \in E} p(e_{ij}=1)\cdot \prod_{(i,j) \notin E} p(e_{ij}=0)$. Such prior uses a uniform probability controlling the sparsity of the graph $G$, and requires prior knowledge of the sparsity of the graph to choose the probability parameter. In this paper, we propose a new prior on $G$ which induces adaptive sparsity in the graphical models as:
\begin{align*}
G & \propto  \exp(-|G|^{\xi}),
\end{align*}
where $\xi$ is a positive value penalizing on the size of the graph $G$. Varying $\xi$ penalizes a graph size with different strength. A large value of $\xi$ $(>3)$ constrains the graph to be extremely sparse, while a value of $\xi$ near zero approximates a uniform prior on all graphs. In the Bayesian method, we assign a uniform prior between 0 and a large value (e.g. 5) on $\xi$, and estimate $\xi$ using an MCMC algorithm. Such choice of prior lets the data choose the intensity of the penalization on the graph size and leads to adaptive sparsity in the inferred graph $G$.

\subsection{Bayesian Hierarchical Model for Unrestricted Graphs} \label{subsec:model3}

In this section, we assume the graphical model of the residuals is unrestricted. The HIW prior was extended for nondecomposable graphical models by Giudici and Green (1999), Roverato (2002), Brooks et al. (2003), and Atay-Kayis and Massam (2005). However, sampling nondecomposable graphs from HIW distributions induces extensive computational burden. In Section \ref{subsec:model12}, we employ a Bayesian graphical lasso selection prior to achieve sparse estimation of $S$. We now apply the Bayesian graphical lasso selection prior on $C$, the inverse of $S$. As mentioned previously, the graphical lasso prior on a matrix $C$ is equivalent to  a joint density of  exponential distributions on the diagonal elements $C_{jj}$, $j=1,\ldots,q$, and Laplace distributions on the off-diagonal elements $C_{jj'}$, $j<j'$. The Bayesian graphical lasso method does not set the off-diagonal elements in $C$ to exact zeros. Hence similarly,  we add a point mass at $0$ in the Laplace priors. The priors are detailed as follows:
\begin{align*}
C | \lambda^C, \boldsymbol\rho^C  \sim & con_{\lambda^C,\rho^C}^{-1} \prod_{j<j'} \Big[ (1-\rho^C_{jj'}) \mathcal{I}\{C_{jj'}=0\} + \rho^C_{jj'} \mathrm{Laplace}(\lambda^C) \Big] \cdot \prod^q_{j=1} \mathrm{Exp}(\frac{\lambda^C}{2}) \cdot   \mathcal{I}\{ C \in \mathcal{C}^+\},
\end{align*}
where $\mathcal{C}^+$ is the space of all positive-definite concentration matrices, and $con_{\lambda^C,\rho^C} = \int_{C \in \mathcal{C}^+}  \prod_{j<j'} \Big[ (1-\rho^C_{jj'}) \mathcal{I}\{C_{jj'}=0\} + \rho^C_{jj'} \mathrm{Laplace}(\lambda^C) \Big] \prod^q_{j=1} \mathrm{Exp}(\frac{\lambda^C}{2}) dC$ is the normalizing constant, which depends on $\lambda^C$ and $\boldsymbol\rho^C = \{\rho^C_{jj'},j<j'\}$.

Again, we choose an extension of the conjugate priors on $\lambda^C$ and $\boldsymbol\rho^C$ to induce computationally tractable full conditional distributions for posterior inference:
\begin{align*}
(\lambda^C, \boldsymbol\rho^C) & \sim  con_{\lambda^C,\rho^C} \mathrm{Gamma}(a_\lambda, b_\lambda) \prod_{j<j'} \mathrm{Beta}(a_\rho, b_\rho).
\end{align*}
As mentioned previously, the hyper-parameter $\lambda^C$ shrinks the entries in $C$ toward zero, while $\rho^C_{jj'}$ controls the probability that the $(j,j')$ element will be enforced to be a zero. Based on our experiments, we find that noninformative priors on $\rho^C_{jj'}$ would lead to significant inaccuracy in estimating $S$, and influence the factor selection and loadings estimation. In this case, we specify $(a_\lambda,b_\lambda)=(1,1)$ for a diffuse prior for $\lambda^C$, and $(a_\rho,b_\rho)=(1,q)$ for a sparse prior of $\rho^C_{jj'}$.

\subsection{Posterior Inference}

We derive the full conditionals for all the parameters and perform the posterior inference using a Gibbs sampling algorithm. Note that the prior specification for the parameters $\{M, z_k, \tau_k\}$ in the graphical factor analysis models parallels the hierarchical model in formulas (\ref{eq:L_hierarchy}), so the full conditionals of these parameters for a Gibbs algorithm are similar to those in Section \ref{subsec:model1_postsample}. In this section, we just present the sampling algorithm of the parameter set $\{S,G,\xi\}$ for a decomposable graph of the residuals, and $\{C,\lambda^C,\boldsymbol\rho^C\}$ for an unrestricted graph of the residuals, in sequence.

\subsubsection{Sampling $\{S,G,\xi\}$ for decomposable graphical models}

\begin{enumerate}
\item Sampling of $S$: The full conditional distribution of $S$ is
    \begin{align*}
    S|G,y,M,f & \sim \mathrm{HIW}(G,\delta+n,\Phi+\Lambda),
    \end{align*}
    where $\Lambda=(y-Mf)(y-Mf)^T$. Hence we can generate posterior samples of $S$ directly from the HIW distribution conditional on other parameters.

\item Sampling of $G$: The conditional distribution of $G$ is
    \begin{align*}
    G|y,M,f,\xi & \propto \frac{h(G,\delta,\Phi)}{h(G,\delta+n-1,\Phi+\Lambda)} \cdot \exp\{-|G|^\xi\}.
    \end{align*}
    The term $h(G,\delta,\Phi)$ is the normalizing constant for the $\mathrm{HIW}(G,\delta,\Phi)$ distribution given by $$ h(G,\delta,\Phi)=\frac{\prod_{k=1}^{K}|\frac{\Phi_{P_k}}{2}|^{(\frac{\delta+|P_k|-1}{2})}\Gamma_{|P_k|} \big( \frac{\delta+|P_k|-1}{2} \big)^{-1}} {\prod_{k=2}^{K}|\frac{\Phi_{Q_k}}{2}|^{(\frac{\delta+|Q_k|-1}{2})}\Gamma_{|Q_k|} \big(\frac{\delta+|Q_k|-1}{2} \big)^{-1}}, $$
    where $\Gamma_p(x)=\pi^{p(p-1)/4}\prod_{j=1}^{p}\Gamma(x+(1-j)/2)$ is the multivariate gamma function. Note that the conditional distribution of $G$ is marginalized over $S$ and hence only dependent on $M$, $f$ and $\xi$. This marginalized posterior conditional of $G$ leads to a collapsed Gibbs algorithm in sampling $G$, accelerating the graphical model search task. To sample the graph $G$ from the conditional distribution, we use a random walk MH algorithm within the Gibbs sampling.

    Let $\{e_{jj'}: j<j'\}$ be the set of edge indicators where $e_{jj'}=1$ if $(j,j') \in E$ and $e_{j,j'}=0$ otherwise. In an iteration with the current state of graph $G^c = (V, E^c)$, we choose a pair $(j,j')$ at random and change the state of the edge, i.e. $e^p_{jj'}=1-e^c_{jj'}$. If the proposed state $G^p=(V, E^p)$ is decomposable, the proposal is accepted as a new state with the probability
    \begin{align*}
    \alpha(G^c,G^p) = \min \left\{1,\frac{p(G^c|y,M^c,f^c,\xi^c)}{p(G^p|y,M^c,f^c,\xi^c)} \right\},
    \end{align*}
    where $p(\cdot)$ denotes the posterior conditional distribution of $G$.  If the proposed state $G^p=(V,E^p)$ is not decomposable, then choose another pair $(j,j')$ until the proposal graph is decomposable.

\item Sampling of $\xi$: The conditional distribution of $\xi$ is
    \begin{align*}
    \xi|G & \propto \frac{\exp\{-|G|^\xi\}}{\sum_{G^*} \exp\{-|G^*|^\xi\}}\mathcal{I}\{\xi \in (0,5)\}.
    \end{align*}
    We use a random-walk MH algorithm to generate posterior samples of $\xi$. Given the current state $\xi^c$, generate a proposal $\log(\xi^p)$ from a normal distribution $\mathcal{N}(\log(\xi^c),\sigma_\xi^2)$, with the standard deviation $\sigma_\xi$ chosen properly. We generate the MCMC samples of $\xi$ in the log scale to ensure positivity. The proposal $\xi^p$ is then accepted with the probability
    \begin{equation*}
    \alpha(\xi^c,\xi^p) = \min \left\{1,\frac{p(\xi^p|G^c)}{p(\xi^c|G^c)} \right\},
    \end{equation*}
    where $p(\cdot)$ denotes the full conditional distribution of $\xi$.

\end{enumerate}

\subsubsection{Sampling $\{C,\lambda^C,\rho^C\}$ for unrestricted graphical models}

For convenience, let $\Lambda=(y-Mf)(y-Mf)^T$.

\begin{enumerate}

\item Sampling of the diagonal elements of $C$, $C_{jj}$, for $j=1,\ldots,q$: The full conditional density of $C_{jj}$ is
        \begin{align*}
        p(C_{jj} | \cdot) \propto & (det C)^{n/2} \exp \Big( -\frac{1}{2}\Lambda_{jj}C_{jj} - \frac{\lambda^C}{2}C_{jj} \Big) \mathcal{I} \{C \in \mathcal{C}^+ \}.
        \end{align*}
    Without loss of generality, suppose that $j=q$. Let $C=R'R$ be the Cholesky decomposition of $C$ where the matrix $R=(R_{jj'})$ is upper triangular. Then
        \begin{align*}
        p(C_{jj} | \cdot) \propto & (C_{jj}-c)^{n/2} \exp \Big\{ -(\frac{\Lambda_{jj}}{2}+\frac{\lambda^C}{2}) C_{jj} \Big\} \mathcal{I} \{C_{jj}>c \},
        \end{align*}
    where $c=\sum^{q-1}_{j=1} R_{j,q}^2$ does not depend on $C_{jj}$, and the indicator function $\mathcal{I} \{C_{jj}>c \}$ ensures $C$ to be positive definite. Let $\nu=(C_{jj}-c)$, then the conditional distribution of $\nu$ is
        \begin{align*}
        p(\nu | \cdot) \propto & \nu^{n/2} \exp \Big\{ -(\frac{\Lambda_{jj}}{2}+\frac{\lambda^C}{2}) \nu \Big\} \mathcal{I} \{\nu>0 \},
        \end{align*}
    which follows $\mathrm{Gamma}(\frac{n}{2}+1,\frac{\Lambda_{jj}+\lambda^C}{2})$. Hence, we can draw samples of $\nu$ from the Gamma distribution first, and obtain $C_{jj}=\nu+c$.

\item Sampling of the off-diagonal elements of $C$, $C_{jj'}$, for $j<j'$:  The complete conditional density of $C_{jj'}$ is
        \begin{align*}
        p(C_{jj'} | \cdot) \propto & (det C)^{n/2} \exp ( -\Lambda_{jj'}C_{jj'} ) p(C_{jj'} |\rho^C_{jj'},\lambda^C) \mathcal{I} \{C \in \mathcal{C}^+ \}.
        \end{align*}
    Without loss of generality, suppose that $j=q-1$ and $j'=q$. Then using Lemma 2 of Wong et al.(2003),
        \begin{align*}
        p(C_{jj'} | \cdot) \propto & \Big\{ 1-\frac{(C_{jj'}-a)^2}{cb^2} \Big\}^{n/2} \exp ( -\Lambda_{jj'}C_{jj'})  p(C_{jj'} |\rho^C_{jj'},\lambda^C) \mathcal{I} \{|C_{jj'}-a|<b\sqrt{c} \},\\
            \propto & \mathcal{I} \{|C_{jj'}-a|<b\sqrt{c} \} \cdot \left[ (1-\rho^C_{jj'}) \Big\{ 1-\frac{(C_{jj'}-a)^2}{cb^2} \Big\}^{n/2} \mathcal{I} \{C_{jj'}=0 \}  \right. \\
            & \left.  + \rho^C_{jj'} \Big\{ 1-\frac{(C_{jj'}-a)^2}{cb^2} \Big\}^{n/2} \exp\{-\Lambda_{jj'}C_{jj'}-\lambda^C |C_{jj'}|\} \mathcal{I} \{C_{jj'} \neq 0 \} \right] ,
        \end{align*}
    where $a=\sum^{q-2}_{j=1} R_{j,q-1}R_{j,q}$, $b=R_{q-1,q-1}$, and $c=R_{q-1,q}^2+R_{q,q}^2$ do not depend on $C_{jj'}$. Now transform $C_{jj'}$ to $\nu= (C_{jj'}-a)/(b\sqrt{c})$, and let $\kappa=-a/(b\sqrt{c})$. The full conditional density of $\nu$ is
        \begin{align*}
        p(\nu|\cdot) & \propto (1-\rho^C_{jj'})(1-\kappa^2)^{n/2} \mathcal{I} \{\nu=\kappa,|\nu| < 1\} + \frac{\rho^C_{jj'} \lambda^C}{2} g(\nu),
        \end{align*}
    where $g(\nu)=(1-\nu^2)^{n/2} \exp \{ -\Lambda_{jj'}(\nu  b\sqrt{c} +a) -\lambda^C|\nu  b\sqrt{c} +a| \} \mathcal{I} \{|\nu|<1\}$. The positive definite constraint on $C$ is ensured by the indicator function $ \mathcal{I} \{|\nu|<1\}$. The continuous part of the conditional distribution of $\nu$, $g(\nu)$, cannot be sampled directly. Since $g(\nu)$ only has density over $(-1,1)$ and is zero elsewhere, we can use an independent MH algorithm as the sampling algorithm for $S_{jj'}$ in Section \ref{subsec:model1_postsample}. The details of the independent MH algorithm are explained in Section \ref{subsec:model1_postsample}. Briefly, we choose $\kappa-1$ equally spaced grids between (-1,1), $-1=\nu_0<\nu_1<\cdots<\nu_{\kappa}=1$, which divide the domain of $\nu$ into $\kappa$ intervals of width $2/\kappa$, and construct a piecewise uniform distribution $g_a(\nu)$ approximating $g(\nu)$. The independent MH proposal for generating $\nu$ is then given by
    \begin{align*}
    q(\nu|\cdot) \propto & (1-\rho^C_{jj'}) (1-\kappa^2)^{n/2} \mathcal{I} \{\nu=\kappa,|\nu| < 1\}  + \frac{\rho^C_{jj'} \lambda^C}{2} g_a(\nu).
    \end{align*}
    Samples of $\nu$ could be generated using an inverse-CDF method from the proposal distribution, and the proposal $\nu^*$ is accepted with the probability
        \begin{align*}
        \alpha = \min\left\{ 1, \frac{p(\nu^*)}{q(\nu^*)} \Big/ \frac{p(\nu^c)}{q(\nu^c)} \right\},
        \end{align*}
    where $\nu^c = (C^c_{jj'}-a)/(b\sqrt{c})$ denotes the current state of $\nu$. Samples of $C_{jj'}$ are obtained as $C_{jj'}=\nu b\sqrt{c} +a$.

\item Sampling the shrinkage parameter $\lambda^C$:

The full conditional of $\lambda$ is
\begin{align*}
\lambda^C|C \sim \mathrm{Gamma}(a_\lambda+m, b_\lambda+\sum_{j<j'}|C_{jj'}|+\frac{1}{2}\sum_{j=1}^q C_{jj}),
\end{align*}
where $m$ equals to size of the set $\{(j,j'): j \leq j', C_{jj'} \neq 0 \}$.

\item Sampling the selection parameters $\rho^C_{jj'}$:

The full conditional of $\rho^C_{jj'}$ for $j<j'$ is
\begin{align*}
\rho^C_{jj'} | C \sim \mathrm{Beta}(a_\rho+\mathcal{I} \{C_{jj'} \neq 0\}, b_\rho+\mathcal{I} \{ C_{jj'}=0 \}).
\end{align*}

\end{enumerate}

\subsection{Simulation} \label{subsec:simu2}

In this section, we examined the graphical factor models using simulated datasets. We considered three graphical factor models to generate datasets.

\begin{figure}
\centering
\centerline{\includegraphics[width=6in]{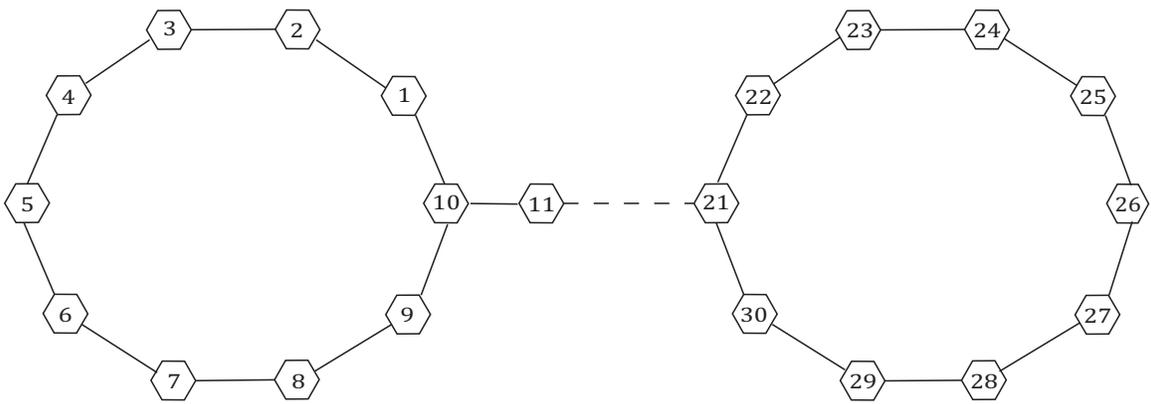}}
\caption{Graphical structure of the residuals in model 6 in the simulations in Section~\ref{subsec:simu2}}\label{figure4}
\end{figure}

\begin{itemize}
\item Model 4: An initial simulation study considered a one-factor model for a 30-dimensional random vector. The factor loading vector $M$ was randomly generated with $\parallel \mathbf{M} \parallel = 1$, the variance of the factor was set to be $\tau^2=4$, and the covariance matrix of the residuals $S$ corresponded to an AR(1) model with $S_{jj'}=0.7^{|j-j'|}$.

\item Model 5: The second simulation considered a two-factor model for a 30-dimensional random vector. The $q \times 2$ factors loadings matrix $M=(\mathbf{M}_1,\mathbf{M}_2)$ with $|\{j: M_{1j} \neq 0\}|=q/2$, $\{j: M_{2j} \neq 0\}= \{1,\ldots,q\} \backslash \{j': M_{1j'} \neq 0\}$, and $\parallel \mathbf{M}_1 \parallel = \parallel \mathbf{M}_2 \parallel = 1$. The variances of the factors were $\boldsymbol\tau^2=(4,4)$, and the covariance matrix of the residuals $S$ was a block diagonal matrix with each square block matrix $B$ of dimension 5, and $B=0.7\mathbf{11}^T+0.3I$.

\item Model 6: The third simulation considered a one-factor model for a 30-dimensional random vector. The factor loading vector $\mathbf{M}$ was randomly generated with $\parallel M \parallel = 1$, the variance of the factor was set to be $\boldsymbol\tau^2=4$, and the covariance matrix of the residuals $S$ corresponds to an nondecomposable graphical model depicted in Figure~\ref{figure4}.

\end{itemize}

We generated datasets with varying sample sizes $n=100$, and $300$ for each model. The proposed HIW based graphical factor model for decomposable graphs was used to recover the number of factors and the graph in model 4 and 5, and the lasso based graphical factor model for unrestricted graphs was used for model 6. The estimates  were based on the posterior samples of 10,000 iterations after 5,000 burn-in iterations. The number of latent factors was estimated by the mode of the posterior distribution, and the graphical model was based on the FDR-based model selection method with an overall FDR of 0.20.

Table \ref{table3} summarizes the results in estimating the number of factors and the graphical model of the residuals over 20 replications. The results indicate that our Bayesian graphical factor models can recover the number of true latent factors most of the time. Besides, it can recover the graphical model of the residuals with both low rates of false positives and false negatives when the sample size is sufficient relative to the dimension.


\subsection{Real Data Analysis} \label{subsec:real2}

We also applied the graphical factor models to a microarray gene expression dataset from Liu et al. (2011), which was obtained from 176 primary breast cancer patients. We focused on 15 mRNA transcripts whose coding genes are known to lie in the estrogen receptor (ER) pathway. The estrogen pathway regulates a variety of genes and plays key roles in the development or progression of breast cancer.

\begin{figure}
\centering
\centerline{\includegraphics[width=6in]{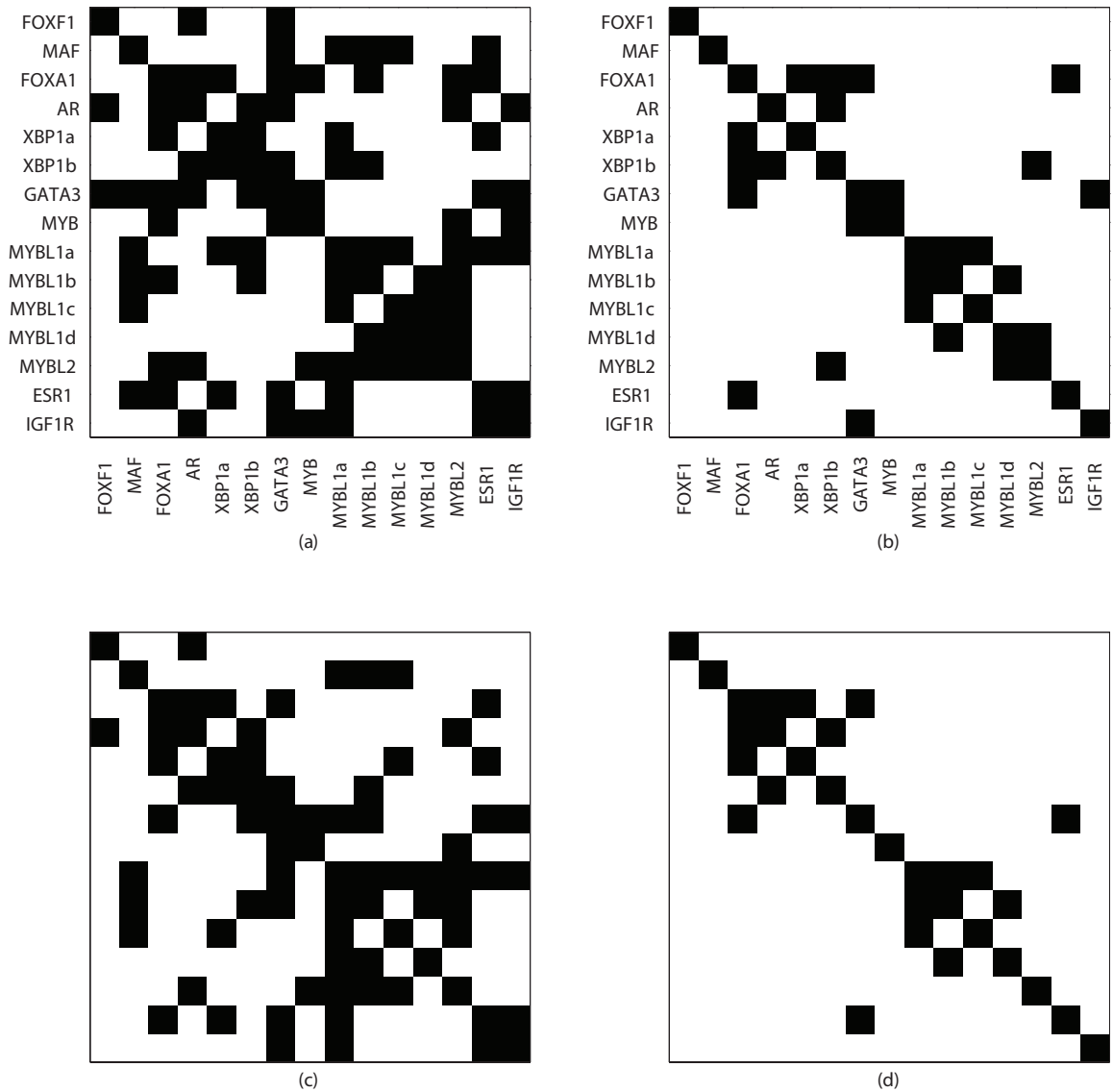}}
\caption{Top panel: The adjacency matrix of the genes involved in ER pathway depicting the graphical model of the residuals inferred by (a) the HIW based graphical factor model (b) the lasso based graphical factor model.  Bottom panel: The adjacency matrix depicting the edges that are selected for over half of the simulation runs by (c) the HIW based graphical factor model and (d) the lasso based graphical factor model, respectively. The simulated datasets were generated from a graphical factor model where the parameters were  the posterior mean estimates of the real data analysis by the lasso based graphical factor model. }\label{figure5}
\end{figure}

We analyzed the data with the two graphical factor models.  The HIW based graphical factor model identifies one latent factors, and the lasso based model does not detect any latent factors. The adjacency matrices corresponding to the inferred graphical models by the HIW and lasso based methods are plotted in Figure~\ref{figure5} (a) and (b) respectively, depicting the conditional dependency relationship among the variables. Some of the genes have multiple sets of oligonuleotide sequences on the microarray, and hence have multiple appearances including MYBL1 (MYBL1a, MYBL1b, MYBL1c, MYBLd) and XBP (XBP1a, XBP1b). The plots of the adjacency matrices show that the HIW based method derived a graph sparser than the graph inferred by the graphical lasso selection method, with the direct regulatory relationships identified in (a) mostly included in (b) as well. The difference in the sparsity level of the graphs and the inferred number of latent factors could be due to the restriction of decomposition in the HIW method.


We conducted a further simulation study based on the results of the above real data analysis. We treated the means of the posterior distributions of the parameters by the lasso based method as the true values, generated simulated data with the same dimension and sample size, i.e. $q=15$ and $n=176$, and then reapplied the two graphical factor models to the simulated datasets. We conducted 20 runs of the simulation based on the real data analysis. The adjacency matrices in Figure~\ref{figure5} (c) and (d) show the edges that are selected for over half of the runs by the HIW based and the lasso based graphical factor model respectively. We note that the simulation results are consistent with the real data analysis in that the graphical model identified by the HIW based model is much sparser than the lasso based model, and the inferred conditional associations in the real analysis are mostly re-identified in the simulations.

\section{Discussion}  \label{sec:diss}

In this paper, we propose a Bayesian method for estimating covariance matrices of a particular structure, which is a summation of a low rank and a sparse component. Different from the frequentist LOREC method, which is based on the sample covariance estimate only, our Bayesian method of covariance decomposition is likelihood-based. Hence, it takes the variability of the variables into consideration in case of $p>>n$ and shrinks the covariance elements of varying intensities differently, as indicated in the real data analysis. Furthermore, in stead of the point estimates of the parameters as in the frequentist LOREC, the Bayesian covariance decomposition method yields posterior distributions of the parameters providing a degree of uncertainties in model inference. We model the low rank and sparse component in the form of a latent factor model with correlated residuals. The representation of the decomposable covariance facilitates a Bayesian inference by using conjugate priors on all the parameters except for the off-diagonal elements in the sparse component as well as providing estimates of factor loadings in addition to the estimate of the low rank and sparse component.  Simulations indicate that such representation favors the covariance estimation for a latent factor model but does not perform as good as the frequentist LOREC method for a random effect model.

We further extend our method to a graphical factor model, in which we perform inference on both the number of latent factors and the sparse graphical model of the residuals. Simulation studies show that the methods can successfully recover the number of factors as well as the graphical model when the sample size is sufficient relative to the dimension. However, simulations (not presented here) also indicate that in the case of $p >> n$, the methods tend to choose over-sparse graphical models of the residuals. This is reasonable: when the sample size is small, the estimate of the covariance of the residuals $S$ would be inaccurate resulting in significant change in estimating the $S$ inverse. In the high-dimensional low-sample-size condition, the Bayesian methods would choose the most sparse estimate of $S$ inverse that fits the data.

In the graphical factor analysis, we use the HIW method to model a decomposable graph of the residuals and the Bayesian graphical lasso method for an unrestricted graph. One way to determine which method fits the data better in application is to calculate the marginal likelihoods of the inferred models based on the MCMC outputs as discussed in Chib (1995) and Chib and Jeliazkov (2001). Briefly, as the marginal likelihood of a model $M$ is the normalizing constant of the posterior density, we can obtain the logarithm of the marginal likelihood as 
$$\log m(y|M) = \log f(\mathbf{y}|M,\boldsymbol\theta^*) + \log \pi(\boldsymbol\theta^*|M) - \log \pi(\boldsymbol\theta^*|\mathbf{y},M), $$
where $\boldsymbol\theta^*$ is some selected point, such as the posterior mode, of the parameter vector $\boldsymbol\theta$ for model $M$. Suppose that the parameter vector could be split into B blocks $\boldsymbol\theta^*=(\boldsymbol\theta^*_1,\ldots,\boldsymbol\theta^*_B)$. Then, the posterior density at the $\boldsymbol\theta^*$ could be written as
$$\pi(\boldsymbol\theta^*|\mathbf{y},M) = \pi(\boldsymbol\theta^*_1|\mathbf{y},M)\pi(\boldsymbol\theta^*_2|\mathbf{y},M,\boldsymbol\theta^*_1)\cdots \pi(\boldsymbol\theta^*_B|\mathbf{y},M,\boldsymbol\theta*_1,\ldots,\boldsymbol\theta^*_{B-1}), $$
where the first term is the marginal ordinate that can be estimated by $\boldsymbol\theta_1$'s posterior distribution from the initial Gibbs samples, and the remaining terms are reduced conditional ordinates. Each reduced conditional ordinate, $\pi(\boldsymbol\theta^*_t|\mathbf{y},M,\boldsymbol\theta^*_1,\ldots,\boldsymbol\theta^*_{t-1})$, can be estimated by $\boldsymbol\theta_t$'s posterior distribution from a reduced Gibbs MCMC run in which $\boldsymbol\theta_1,\ldots,\boldsymbol\theta_{t-1}$ are fixed at $\{\boldsymbol\theta^*_1,\ldots,\boldsymbol\theta^*_{t-1}\}$ and sampling is over $\{\boldsymbol\theta_t,\ldots,\boldsymbol\theta_B\}$. Since the full conditional distributions of the parameters  are explicitly given for the HIW and lasso based graphical factor models, it is straightforward to estimate the reduced conditional ordinates by fixing certain parameter blocks and running additional reduced Gibbs iterations without new programming required. Hence, we can subsequently estimate the posterior ordinate $\pi(\boldsymbol\theta^*|\mathbf{y},M)$ and then the marginal likelihood of the inferred models.

\section*{Appendix: Derivation of the Conditional Densities for the Elements in $S$}
\setcounter{section}{1}
\renewcommand{\thesection}{\Alph{section}}
\subsection{The Conditional Densities for the Diagonal Elements in S}

For convenience, let $\Lambda=(y-Mf)(y-Mf)^T$.

For each diagonal element $S_{jj}$ in S, the full conditional density is
\begin{align*}
p(S_{jj} | \cdot) \propto & \det(S)^{-n/2} \exp\{- \frac{1}{2} tr(S^{-1} \Lambda) - \frac{\lambda}{2} S_{jj} \} \mathcal{I} \{S \in \mathcal{S}^+ \}.
\end{align*}
Without loss of generality, suppose that $j=p$. Let $S = \left[ \begin{array}{cc}
                                                                    S_{-j,-j} & S_{-j,j} \\
                                                                    S_{j,-j} & S_{jj} \end{array} \right]$.
With the property of matrices, we have
\begin{align*}
\det(S) = & \det(S_{-j,-j}) \cdot \det(S_{jj}-S_{j,-j}S^{-1}_{-j,-j}S_{-j,j}),\\
             \propto & (S_{jj}-c), \text{ where } c = S_{j,-j}S^{-1}_{-j,-j}S_{-j,j}. \\
\\
S^{-1} = & \left[ \begin{array}{cc}
                     S_{-j}+ S^{-1}_{-j,-j}S_{-j,j}(S_{jj}-c)^{-1}S_{j,-j}S^{-1}_{-j,-j} & -S^{-1}_{-j,-j}S_{-j,j}(S_{jj}-c)^{-1} \\
                     -(S_{jj}-c)^{-1}S_{j,-j}S^{-1}_{-j,-j} & (S_{jj}-c)^{-1} \end{array} \right], \\
            = & \left[ \begin{array}{cc}
                       S^{-1}_{-j,-j} & 0 \\
                       0 & 0 \end{array} \right] + \left[ \begin{array}{c}
                                                        S^{-1}_{-j,-j}S_{-j,j} \\
                                                        -1 \end{array} \right] (S_{jj}-c)^{-1} \left[ \begin{array}{cc} S_{j,-j}S^{-1}_{-j,-j} & -1 \end{array} \right]. \\
\\
tr(S^{-1}\Lambda) = & tr \Big( \left[ \begin{array}{cc}
                       S^{-1}_{-j,-j} & 0 \\
                       0 & 0 \end{array} \right] \Lambda \Big)  + tr \Big(\left[ \begin{array}{c}
                                                        S^{-1}_{-j,-j}S_{-j,j} \\
                                                        -1 \end{array} \right] (S_{jj}-c)^{-1} \left[ \begin{array}{cc} S_{j,-j}S^{-1}_{-j,-j} & -1 \end{array} \right] \Lambda \Big) ,\\
                 = & tr \Big( \left[ \begin{array}{cc}
                       S^{-1}_{-j,-j} & 0 \\
                       0 & 0 \end{array} \right] \Lambda \Big)  +  d(S_{jj}-c)^{-1}  ,\\
                       & \text{ where } d = \left[ \begin{array}{cc} S_{j,-j}S^{-1}_{-j} & -1 \end{array} \right] \Lambda \left[ \begin{array}{c} S^{-1}_{-j,-j}S_{-j,j} \\  -1 \end{array} \right].
\end{align*}

Hence, we have
\begin{align*}
p(S_{jj} | \cdot) \propto (S_{jj}-c)^{-n/2} \exp \{-\frac{d}{2}(S_{jj}-c)^{-1} -\frac{\lambda}{2} S_{jj} \} \mathcal{I} \{S_{jj}>c\}.
\end{align*}

\subsection{The Conditional Densities for the Off-diagonal Elements in S}

The conditional density of the off-diagonal element of S, $S_{jj'}$, is
\begin{align*}
p(S_{jj'}|\cdot) \propto & \det(S)^{-n/2} \exp\{-\frac{1}{2}tr(S^{-1}\Lambda)\} \Big[ (1-\rho_{jj'}) \mathcal{I}\{S_{jj'}=0\} + \rho_{jj'} \frac{\lambda}{2} \exp(-\lambda |S_{jj'}|) \Big] \mathcal{I} \{S \in \mathcal{S}^+ \}.
\end{align*}
Without loss of generality, suppose that $j=p-1$ and $j'=p$. Let $S = \left[ \begin{array}{cc}
                                                                    S_{-(jj')} & S_{-(jj'),jj'} \\
                                                                    S_{jj',-(jj')} & S_{jj',jj'} \end{array} \right]$,
where $S_{jj',jj'}=\left[ \begin{array}{cc} S_{jj} & S_{jj'} \\ S_{jj'} & S_{j'j'} \end{array} \right]$.

With the property of matrices, we have
\begin{align*}
\det(S) \propto & \det(S_{jj',jj'}-B), \text{ where } B=S_{jj',-(jj')}S^{-1}_{-(jj')}S_{-(jj'),jj'}, \\
\\
S^{-1} = & \left[ \begin{array}{cc} S_{-(jj')}^{-1} & 0 \\ 0 & 0 \end{array} \right] + \left[ \begin{array}{c}
                    S_{-(jj')}^{-1}S_{-(jj'),jj'} \\ -I_2 \end{array} \right] (S_{jj'}-B)^{-1} \left[ \begin{array} {cc} S_{jj',-(jj')}S_{-(jj')}^{-1} & -I_2 \end{array} \right], \\
\\
tr(S^{-1}\Lambda) = & tr(\left[ \begin{array}{cc} S_{-(jj')}^{-1} & 0 \\ 0 & 0 \end{array} \right]S) + tr( (S_{jj',jj'}-B)^{-1}D),\\
                   & \text{  where } D = \left[ \begin{array} {cc} S_{jj',-(jj')}S_{-(jj')}^{-1} & -I_2 \end{array} \right] \Lambda \left[ \begin{array}{c} S_{-(jj')}^{-1}S_{-(jj'),jj'} \\ -I_2 \end{array} \right].
\end{align*}

Hence, we have
\begin{align*}
p(S_{jj'}|\cdot) \propto  &  \det(S_{jj',jj'}-B)^{-n/2} \exp \{-\frac{1}{2}tr((S_{jj'}-B)^{-1}D)\}
                             \times p(S_{ij}|\lambda,\rho_{jj'}) \mathcal{I}\{S \in \mathcal{S}^+\},\\
                 \propto  &  \left\{ 1-\frac{(S_{jj'}-B_{12})^2}{(S_{jj}-B_{11})(S_{j'j'}-B_{22})}\right\}^{-\frac{n}{2}} \\
                          & \cdot \exp \left\{ -\frac{1}{2}\frac{(S_{j'j'}-B_{22})D_{11}+(S_{jj}-B_{11})D_{22}-2(S_{jj'}-B_{12})D_{12})}
                          {(S_{jj}-B_{11})(S_{j'j'}-B_{22})-(S_{jj'}-B_{12})^2}  \right\} \\
              & \cdot \Big[ (1-\rho_{jj'}) \mathcal{I} \{S_{jj'}=0\} + \rho_{jj'} \frac{\lambda}{2} \exp(-\lambda |S_{jj'}|) \Big]  \\
		&	\cdot \mathcal{I} \left\{(S_{jj'}-B_{12})^2<(S_{jj}-B_{11})(S_{j'j'}-B_{22}) \right\}.
\end{align*}

\clearpage\pagebreak\newpage

\begin{landscape}
\begin{table}
\vspace*{-6pt}
\caption{\label{table2}Simulation results for Bayesian low rank and sparse matrix decomposition for model 1, 2 and 3. The mean $L_1$ and Frobenius norms of the difference matrices between the estimated and true covariance matrices over 20 replications are presented in the table with the standard deviations in parentheses. See Section \ref{sec:simu1} for details about the models.}
\centering
\begin{tabular*}{\columnwidth}{@{}l@{\extracolsep{\fill}}l@{\extracolsep{\fill}}c@{\extracolsep{\fill}}c@{\extracolsep{\fill}}c@{\extracolsep{\fill}}c@
{\extracolsep{\fill}}c@{\extracolsep{\fill}}c@{\extracolsep{\fill}}c@{}}
&&&&&&&&\\
\multicolumn{9}{c}{{Losses of Covariance Estimators}} \\
\hline
\hline
\multicolumn{3}{c}{}&\multicolumn{2}{c}{{model 1}}&\multicolumn{2}{c}{{model 2}}&\multicolumn{2}{c}{{model 3}} \\
\cline{4-9}
\multicolumn{3}{l}{q} &{$L_1$ norm}&{Frobenius}&{$L_1$ norm}&{Frobenius}&{$L_1$ norm}&{Frobenius}\\
\cline{1-9}
50 & { } & {Bayesian}     & 11.90 (1.82) & \; 8.46 (0.82) & 12.89 (1.63) & 11.62 (1.46) & 13.40 (2.02) & \; 9.78 (0.70)\\
{} & {} & {LOREC}           & 13.64 (1.76) & \; 9.15 (0.68) & 11.84 (1.77) & \; 9.12 (0.98) & 13.55 (1.73) & \; 9.75 (0.70)\\
{} & {} & {Sample}          & 15.18 (2.12) & 11.63 (0.78) & 13.98 (2.95) & 11.06 (1.27) & 13.69 (1.99) & 10.75 (0.86) \\
100 & { } & {Bayesian}    & 15.28 (2.45) & 10.00 (0.78) & 25.48 (2.74)& 16.13 (1.17)& 20.89 (3.81)& 15.93 (1.00) \\
{} & {} & {LOREC}           & 16.54 (2.19) & 10.37 (0.96) & 23.67 (2.98) & 14.18 (1.68) & 21.97 (3.18) & 15.40 (0.96) \\
{} & {} & {Sample}          & 20.74 (1.90) & 17.36 (0.52) & 26.18 (4.64) & 20.08 (1.80) & 25.39 (5.00) & 19.51 (1.43)\\
200 &{}&{Bayesian}        & 29.90 (3.98) & 18.16 (1.15) & 49.83 (6.28) & 32.71 (9.94) & 35.58 (4.10)  & 25.23 (1.21) \\
{} & {} & {LOREC}           & 32.79 (4.06) & 19.31 (1.77) & 48.96 (7.51) & 28.02 (6.09) & 39.66 (3.37) & 23.68 (1.04) \\
{} & {} & {Sample}          & 42.58 (2.56) & 35.42 (0.97) & 54.49 (4.96) & 37.60 (1.66) & 45.16 (5.25) & 35.76 (1.75) \\
\cline{1-9}
\end{tabular*}
\end{table}
\end{landscape}

\clearpage\pagebreak\newpage

\begin{landscape}

\begin{table}
\vspace*{-6pt}

\caption{\label{table1}Simulation results for Bayesian low rank and sparse matrix decomposition for model 1, 2 and 3. The top panel shows the mean successful rates of rank recovery and the means of the estimated ranks over 20 replications with the standard deviations in parentheses. The bottom panel shows the mean false positives and false negatives in the support recovery of the sparse component $S$. FP: number of false positive discoveries; FN: number of false negative discoveries. See Section~\ref{sec:simu1} for details about the models.}
\centering
\begin{tabular*}{\columnwidth}{@{}l@{\extracolsep{\fill}}l@{\extracolsep{\fill}}c@{\extracolsep{\fill}}c@{\extracolsep{\fill}}c@{\extracolsep{\fill}}c@
{\extracolsep{\fill}}c@{\extracolsep{\fill}}c@{\extracolsep{\fill}}c@{}}
&&&&&&&&\\
\multicolumn{9}{c}{{Rank Recovery}} \\
\hline\hline
\multicolumn{3}{c}{}&\multicolumn{2}{c}{{model 1}}&\multicolumn{2}{c}{{model 2}}&\multicolumn{2}{c}{{model 3}} \\
\cline{4-9}
\multicolumn{3}{l}{q}&{\%(3 factors)}&{mean(se)}&{\%(1 factors)}&{mean(se)}&{\%(3 factors)}&{mean(se)}\\
\cline{1-9}
50&{ }&{ Bayesian}  &  95 & 2.95 (0.22) & 90 & 0.90 (0.31) & 80 & 2.80 (0.41) \\
{}&{}&{ LOREC}        & 100 & 3.00 (0.00)& 100 & 1.00 (0.00) & 40 & 2.05 (1.00) \\
100&{}&{ Bayesian}  & 90 & 2.90 (0.31) & 80 & 1.20 (0.41) & 85 & 2.85 (0.37) \\
{}&{}&{ LOREC}        & 95 & 2.95 (0.22) & 100 & 1.00 (0.00) & 50 & 2.35 (0.93) \\
200&{}&{ Bayesian}  & 100 & 3.00 (0.00) & 90 & 0.92 (0.28) & 90 & 2.90 (0.31) \\
{}&{}&{ LOREC}        & 90 & 2.90 (0.31) & 100 & 1.00 (0.00) & 80 & 2.70 (0.66) \\
\cline{1-9}
\multicolumn{9}{c}{{Sparsity Recovery}} \\
\hline\hline
\multicolumn{3}{c}{}&\multicolumn{2}{c}{{model 1}}&\multicolumn{2}{c}{{model 2}}&\multicolumn{2}{c}{{model 3}} \\
\cline{4-9}
\multicolumn{3}{l}{q} &{FN}&{FP}&{FN}&{FP}&{FN}&{FP}\\
\cline{1-9}
50 &{ }& { Bayesian}    & 0 (0) \! & \; 7.75 (\; 4.06)\;  & \; \; 2.60 (\; 1.90)  \; & \; \; 40.95 (\; 11.38)  \; & \; 13.45 (\; 7.96)   \;  & \; \; 21.70 (\; \; 8.16)\\
{} & {} & { LOREC}          & 0 (0) \! & \; 6.00 (15.75) \;   & \; \; 0.00 (\; 0.00)  \; & \; 188.25 (\; 50.49)     \; & \; \; 6.20 (\; 3.07) \   & \; 518.25 (\; 93.35) \\
100&{}&{ Bayesian}     & 0 (0) \! & 13.70 (\; 6.33) \;  & \; 15.90 (\; 4.28)   \;  & \; \; 68.90 (\; 15.48)    \; & \; 33.40 (\; 6.06)   \;  & \; \; 37.95 (\; 11.33) \\
{}&{}&{ LOREC}              & 0 (0) \! & \; 3.00 (\; 8.05) \; & \; \; 0.05 (\; 0.22)  \; & \; 508.25 (\; 97.80)      \; & \; \; 5.85 (\; 3.10)  \;  & 1011.60 (304.40) \\
200&{}&{ Bayesian}     & 0 (0) \! & \; 2.00 (\; 1.73) \; & 103.00 (11.91)      \;  & \; \; 34.60 (\; 41.13)    \; & 157.00 (15.19)       \;   & \; \; 10.70 (\; \; 3.01) \\
{}&{}&{ LOREC}              & 0 (0) \! & \; 0.80 (\; 1.47) \; & \; \; 0.6 (\; 1.26)    \; & 1175.60 (265.00)         \; & \; \; 6.4 (\; 4.03)   \;  & 1848.30 (495.00) \\
\cline{1-9}
\end{tabular*}
\end{table}

\end{landscape}

\clearpage\pagebreak\newpage

\begin{table}
\vspace*{-6pt}

\caption{\label{table3}Simulation results for Bayesian graphical factor analysis models for model 4, 5 and 6. The top panel shows the mean successful rates of number of factor recovery and the means of the estimated numbers of latent factors over 20 replications with the standard deviations in parentheses. The bottom panel shows the mean false positives and false negatives in the edge selection for the graphical models of the residuals. FP: number of false positive edges; FN: number of false negative edges. See Section~\ref{subsec:simu2} for details about the models.}
\centering
\begin{tabular}{ccccccccc}
&&&&&&&\\
\multicolumn{7}{c}{{Recovery of Number of Factors}} \\
\hline\hline
&\multicolumn{2}{c}{{model 4}}&\multicolumn{2}{c}{{model 5}}&\multicolumn{2}{c}{{model 6}}\\
n& {\%(1 factor)}&{mean(se)} & {\%(2 factor)}&{mean(se)} & {\%(1 factor)}&{mean(se)}\\
\hline
100  & 95 & 1.05 (0.22) & 70 & 2.40 (0.68) & 75 & 1.25 (0.44) \\
300  & 100 & 1.00 (0.00) & 95  & 1.95 (0.22) & 100 & 1.00 (0.00) \\
\hline
\multicolumn{7}{c}{{Graph of Residuals}} \\
\hline\hline
&\multicolumn{2}{c}{{model 4}}&\multicolumn{2}{c}{{model 5}}&\multicolumn{2}{c}{{model 6}}\\
{n} &{FN(se)}&{FP(se)} & {FN(se)}&{FP(se)} & {FN(se)}&{FP(se)}\\
\hline
100  & 0.05 (0.22) & 7.15 (0.49) & 17.35 (3.69) & 6.00 (1.68) & 0.45 (1.05) & 15.80 (2.75) \\
300  & 0.00 (0.00) & 6.90 (0.31) & \; 3.30 (1.59) & 9.00 (1.86) & 0.00 (0.00) & 17.25 (1.80) \\
\hline
\end{tabular}
\end{table}

\clearpage\pagebreak\newpage

\clearpage\pagebreak\newpage

\clearpage\pagebreak\newpage

\clearpage\pagebreak\newpage

\clearpage\pagebreak\newpage

\end{document}